\documentclass[iop]{emulateapj}

\usepackage{natbib,ifthen}
\usepackage{graphicx}
\usepackage{amssymb,latexsym}
\usepackage{url}
\usepackage[latin1]{inputenc}
\usepackage{fp}

\slugcomment{Accepted for publication in \textit{The Astrophysical Journal}, 30 Apr 2012}

\newcommand {\numsamp} {33\,718}
\newcommand {\numvalid} {33\,491}
\newcommand {\numo} {26\,210}
\newcommand {\numc} {7281}
\newcommand {\numx} {1340}
\newcommand {\numxc} {1299}

\newcommand {\numbad} {227}
\newcommand {\numlo} {9442}
\newcommand {\numhi} {16\,768}
\newcommand {\numir} {6089}

\newcommand {\mic} {$\mu$m}
\newcommand {\ben} {\begin{eqnarray}}
\newcommand {\een} {\end{eqnarray}}

\newcommand{\msunperyr}{M$_\odot$ yr$^{-1}$}
\newcommand{\ks}{$K_{\rm s}$}
\newcommand{\jmag}{{\it J}}
\newcommand{\hmag}{{\it H}}
\newcommand{\mlr}{\.M$_{d}$}
\newcommand{\teff}{T$_{\rm eff}$}
\newcommand{\actaa}{Acta Astron.}

\def\2dust{%
             \leavevmode
             $\raisebox{0.3ex}{\mbox{{\sc 2-D}}}\kern-0.55em%
             \raisebox{-0.3ex}{\mbox{{\sc Dust}}}$}
\defcitealias{Srinivasan2009}{Paper I}
\defcitealias{Srinivasan2010}{Paper III}
\defcitealias{Srinivasan2011}{Paper V}
\defcitealias{Sargent2010}{Paper II}
\defcitealias{Sargent2011}{Paper IV}

\begin{document}

\title{The Mass-Loss Return From Evolved Stars to The Large Magellanic Cloud VI: Luminosities and Mass-Loss Rates on Population Scales}

\author{D. Riebel} 
\affil{Department of Physics and Astronomy, The Johns Hopkins University, 3400 North Charles St. Baltimore, MD 21218, USA}
\email{driebel@pha.jhu.edu}

\author{S. Srinivasan}
\affil{UPMC-CNRS UMR7095, Institut d'Astrophysique de Paris, F-75014 Paris, France}

\author{B. Sargent}
\affil{Center for Imaging Science and Laboratory for Multiwavelength Astrophysics, Rochester Institute of Technology, 54 Lomb Memorial Drive, Rochester, NY 14623, USA}
\affil{Space Telescope Science Institute, 3700 San Martin Drive, Baltimore, MD 21218, USA}

\author{M. Meixner}
\affil{Space Telescope Science Institute, 3700 San Martin Drive, Baltimore, MD 21218, USA}
\affil{Department of Physics and Astronomy, The Johns Hopkins University, 3400 N. Charles St. Baltimore, MD 21218, USA}

\shortauthors{Riebel {\it et al.}}
\shorttitle{}

\begin{abstract}

We present results from the first application of the \textit{G}rid of \textit{R}ed Supergiant and \textit{A}symptotic Giant Branch \textit{M}odel\textit{S} (GRAMS) model grid to the entire evolved stellar population of the Large Magellanic Cloud (LMC).  GRAMS is a pre-computed grid of 80\,843 radiative transfer (RT) models of evolved stars and circumstellar dust shells composed of either silicate or carbonaceous dust.  We fit GRAMS models to $\sim$30\,000 Asymptotic Giant Branch (AGB) and Red Supergiant (RSG) stars in the LMC, using 12 bands of photometry from the optical to the mid-infrared.  Our published dataset consists of thousands of evolved stars with individually determined evolutionary parameters such as luminosity and mass-loss rate.  The GRAMS grid has a greater than 80\% accuracy rate  discriminating between Oxygen- and Carbon-rich chemistry.  The global dust injection rate to the interstellar medium (ISM) of the LMC from RSGs and AGB stars is on the order of $1.5\times 10^{-5}$~\msunperyr, equivalent to a total mass injection rate (including the gas) into the ISM of $\sim5\times10^{-3}$~\msunperyr.  Carbon stars inject two and a half times as much dust into the ISM as do O-rich AGB stars, but the same amount of mass.  We determine a bolometric correction factor for C-rich AGB stars in the \ks\ band as a function of \jmag\ -- \ks\ color, BC$_{K_{s}} = -0.40(J-K_{s})^2 + 1.83(J-K_{s}) + 1.29$.  We determine several IR color proxies for the dust mass-loss rate (\mlr) from C-rich AGB stars, such as $\log \dot{M_{d}} = \frac{-18.90}{(K_{s}-[8.0])+3.37}-5.93$.  We find that a larger fraction of AGB stars exhibiting the `long-secondary period' phenomenon are O-rich than stars dominated by radial pulsations, and AGB stars without detectable mass-loss do not appear on either the first-overtone or fundamental-mode pulsation sequences.

\end{abstract}

\section{INTRODUCTION}

At the end of their lives, stars of approximately solar-mass (0.8--8~M$_{\odot}$) ascend the Asymptotic Giant Branch (AGB), the final phase of nuclear burning in the lives of these stars.  The AGB is one of the brightest populations in the infrared (IR) sky, contributing up to $\sim$20--30\% of the IR light for the Small Magellanic Cloud (SMC) \citep{Boyer2011}.  Essentially all AGB stars are variables \citep{Vijh2009},  with periods on the order of hundreds of days.  This variability is caused by hydrodynamic pulsations traveling through the extended atmosphere of the star.  The dramatic changes in stellar radius these shocks produce cause brightness fluctuations on the scale of $\sim$2~mag \citep[e.g.][]{Wood1999}, on time scales of hundreds of days \citep{Fraser2008,Whitelock2008,Riebel2010}.  On much longer time scales ($\sim$10$^{5}$~yr), runaway thermonuclear reactions in the helium burning shell of an AGB star, called `thermal pulses,' can cause much more dramatic brightness variations \citep{Schwarzschild1965, Vassiliadis1993}.  Thermal pulses cause readjustments in the global structure of the AGB star, carrying elements synthesized in the nuclear burning regions to the surface.  This process, known as the `Third Dredge Up,' is responsible for the formation of carbon-rich AGB stars \citep{Iben1983}, as well as bringing more exotic elements such as Tc to the surface \citep{Uttenthaler2010}, and its precise details remain a topic of current research \citep[e.g.][]{Karakas2010}.  Towards the end of their evolution, AGB stars exhibit extensive rates of mass-loss \citep{Wachter2002}, driven by the aforementioned hydrodynamic pulsations and radiation pressure on the resultant dust grains \citep{Winters2000,Mattsson2011}.  This mass-loss, enhanced by elements produced in the nuclear-burning regions of the star and dredged to the surface, makes AGB stars one of the primary sources for $\alpha$-elements in the universe, and an important contributor to dust in galaxies.  The precise degree to which AGB stars contribute dust to the Interstellar Medium (ISM) is a matter of some dispute however.  Specifically, to what extent do AGB stars or supernovae (SNe) dominate this process?

The AGB contribution to a galaxy's total dust budget can be computed if the entire population of mass-losing AGB stars in the galaxy is identified, and the rate of dust production by each star in this sample is known. A comprehensive study of Galactic AGB stars is hampered by extinction in the plane of the Milky Way. The Large Magellanic Cloud (LMC) offers the ideal combination of relative proximity and low line-of-sight extinction, allowing for detailed studies of galaxy-wide evolved star populations.  The LMC's high galactic latitude minimizes both foreground contamination by Milky Way stars and reddening due to intervening dust.  In addition, the distance to the LMC ($\sim$50~kpc) is well determined \citep[e.g.][]{Ngeow2008}.  See \citet{Schaefer2008} for discussion and meta-analysis of this measurement.  This is close enough that individual stars can be resolved, yet far enough away that the 3-D structure of the LMC can be neglected, accurate distances assumed for all stars, and therefore intrinsic brightnesses determined.

Recent large-scale photometric surveys of the LMC such as the \textit{Magellanic Clouds Photometric Survey} \citep[MCPS;][]{Zaritsky2004}, the \textit{Two Micron All-Sky Survey} \citep[2MASS;][]{Skrutskie2006}, and the \textit{Surveying the Agents of a Galaxy's Evolution} survey \citep[SAGE;][]{Meixner2006}, have allowed the construction of catalogs of the entire AGB population of that galaxy, with multi-band photometry for tens of thousands of sources \citep{Blum2006}.  The SAGE-Spec follow-on to the SAGE survey \citet{Kemper2010} has produced spectral classifications for 100 sources in the LMC \citep{Woods2011}.  The IRSF survey \citep{Kato2007} also examined the LMC in near-IR (JH\ks) bands.

The AKARI mission \citep{Murakami2007} has also been used to survey a portion of the LMC \citep{Ita2008}.  While this survey covers a smaller area, the unique [11] and [15]~\mic\ bands of the AKARI satellite are a valuable contribution to the multi-wavelength coverage of this important galaxy.  The ongoing Vista Magellanic Clouds Survey (VMC) \citet{Cioni2011} will push three magnitudes deeper than the AKARI survey, and has already been used to develop a spatially resolved star formation history of part of the LMC \citep{Rubele2011}.

As a first step towards the direct measurement of mass-loss rates for the entire LMC AGB sample, \citep[][hereafter Paper I]{Srinivasan2009} computed infrared excesses from SAGE data and used these to estimate the total dust injection rate. Boyer et al. (2012, in press) have performed a similar analysis for the Small Magellanic Cloud (SMC).  \citet{Matsuura2009} determined a simple IR color proxy for mass-loss using mass-loss rates derived by \citet{Groenewegen2007} and \citet{Gruendl2008} through detailed modeling of individual sources, and extrapolated a global gas and dust budget for the LMC.  An alternative to detailed modeling is to compare the observed SEDs of sources to pre-computed models.  With this aim, \citet[][hereafter Paper IV]{Sargent2011} and \citet[][hereafter Paper V]{Srinivasan2011} presented the Grid of RSG and AGB ModelS (GRAMS) for oxygen-rich and carbonaceous dust respectively.  In this paper we apply the GRAMS to the whole population of evolved stars in the LMC in order to derive more precise measurements of individual sources and the population wide return of mass to the galaxy.

The remainder of this paper is organized as follows: Section~\ref{sec:data} of this paper discusses the sources of our observational data (\S~\ref{sec:dat_source}) and our fitting procedure (\S~\ref{sec:var} \&~\ref{sec:fitting}).  Section~\ref{sec:tour} displays some specific fits representative of our sample.  Section~\ref{sec:discuss} describes the major results of our fitting, including the O-rich/C-rich determination (\S~\ref{sec:oc}), the revealed luminosity function of the evolved stellar population of the LMC (\S~\ref{sec:lum_func}), the integrated dust mass return to the ISM (\S~\ref{sec:mlr} \&~\ref{sec:int_mlr}), and the development of simple observational proxies for dust mass-loss rate (\S~\ref{sec:proxy}).  We review our conclusions in section~\ref{sec:conclusions}.

\section{DATA AND FITTING PROCEDURES} \label{sec:data}
	\subsection{Data} \label{sec:dat_source}
The original SAGE survey was conducted in two epochs, spaced $\sim$3 months apart \citep{Meixner2006}.  The observations from these epochs have been combined into a single mosaic photometry archive and catalog, which is deeper and has smaller photometric errors than the individual epochs.  Our dataset consists of \numsamp\ sources extracted from the SAGE Mosaic Photometry Archive.  These sources have been matched to optical data from the Magellanic Clouds Photometric Survey (MCPS) \citep{Zaritsky2004}, near-infrared photometry from the 2MASS survey \citep{Skrutskie2006}, and the variability information of the MACHO survey \citep{Alcock1999}, allowing us to construct 12 band spectral energy distributions (SEDs) from the U band to 24~\mic\ for most of our sources.  A 2\arcsec\ matching radius was used for all catalog joins.  Details of the matching to the MCPS catalog can be found in the SAGE Data Delivery Document\footnote{\url{http://data.spitzer.caltech.edu/popular/sage/20090922\_en\-hanced/documents/SAGEDataProductsDescription\_Sep09.pdf}}.  The join to the MACHO catalog is discussed more thoroughly in \citet{Riebel2010}.  All non-SAGE photometry has been de-reddened, with the de-reddened flux, $F_{0}$, related to the observed photometry, $F_{\rm obs}$ by $F_{0} = F_{\rm obs} \times 10^{(0.4A_{\lambda})}$.  Our de-reddening coefficients, $A_{\lambda}$, are listed in Table~\ref{tab:dered}.

\begin{deluxetable}{ll}
\tabletypesize{\scriptsize}
\tablecaption{De-reddening Coefficients}
\tablecomments{De-reddening coefficients used in this work for non-SAGE photometry.  SAGE photometry was not de-reddened, as interstellar reddening is negligible at those wavelengths.}
\tablehead{\colhead{Band} & \colhead{$A_{\lambda}$ (mag)}}
\startdata
U & 0.5900 \\
B & 0.5315 \\
V & 0.4590 \\
I & 0.2708 \\
J & 0.1125 \\
H & 0.0652 \\
\ks\ & 0.0372
\enddata
\label{tab:dered}
\end{deluxetable}

\begin{figure*}
\begin{center}
\begin{tabular}{cc}
\rotatebox{0}{\includegraphics*[scale=0.35]{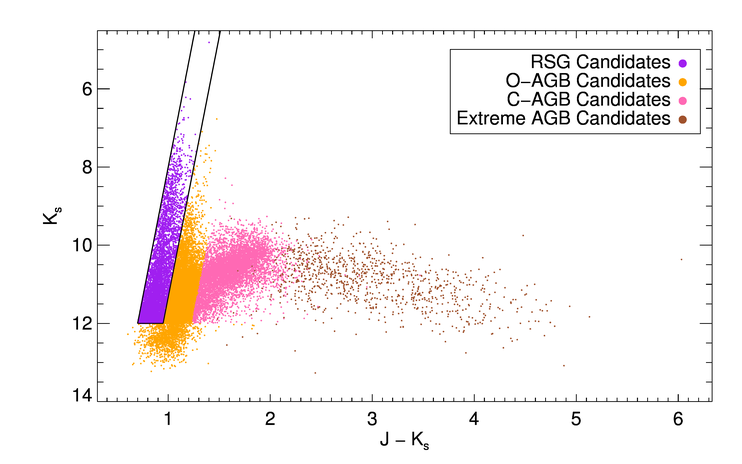}} & \rotatebox{0}{\includegraphics*[scale=0.35]{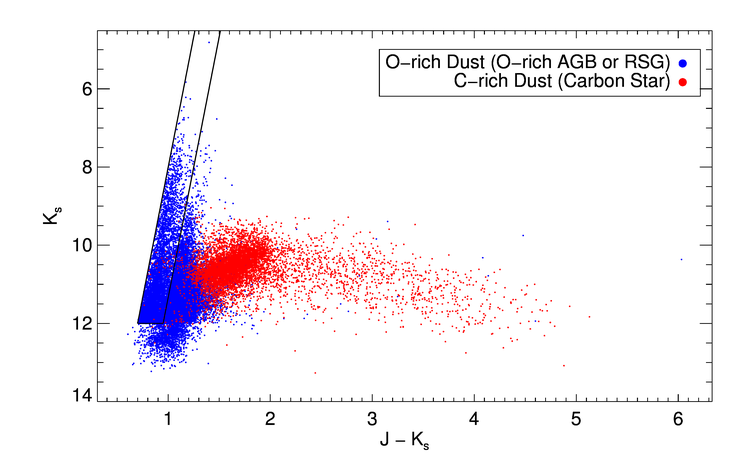}}
\end{tabular}
\caption{\ks\ vs. \jmag\ $-$ \ks\ CMD illustrating the two classification schemes used in this paper.  Color-magnitude cuts \citep[see \S~2.2,][for details]{Riebel2010} are first used to select AGB and RSG candidates (left panel).  The observed SEDs of these candidates are then fit with GRAMS models, which point to either oxygen-rich or carbonaceous dust chemistry (right panel). As the GRAMS chemical classification does not distinguish between O-rich AGB stars and RSGs, we use the CMD cuts of \citet{Boyer2011} (thick lines in both panels) to identify RSG candidates.  Sources with O-rich dust (blue dots) that fall within the CMD cuts are classified as RSGs.}
\label{fig:cmd_class}
\end{center}
\end{figure*}

When extracting sources from the SAGE database, our initial AGB classifications follow from \citet{Cioni2006} and \citet{Blum2006}. A star is classified as an oxygen-rich (O-rich) or carbon-rich (C-rich) AGB candidate based on its location on the \ks\ vs. \jmag\ $-$ \ks\ color magnitude diagram (CMD). The \jmag\ $-$ [3.6] color (or, in the absence of  a J-band detection, the [3.6] $-$ [8.0] color) is used to select extreme AGB candidates. See \S~2.2 in \citet{Riebel2010} for explicit definitions of these color cuts.  RSG candidates are selected based on the color-magnitude criteria presented in \citet{Boyer2011}.  Specifically, RSGs candidates are defined as being brighter than the Tip of the Red Giant Branch (TRGB), $K_{s}=12$, and between the lines $K_s = -13.333(J-K_s +0.25)+24.66$ and $K_s = -13.333(J-K_s)+24.66$.  These CMD-based definitions are illustrated in Figure~\ref{fig:cmd_class}.

Looking for Young Stellar Objects (YSOs), \citet{Gruendl2008} published a list of 13 `Extremely Red Objects' that they spectroscopically identified as carbon stars, based on the presence of SiC absorption.  Twelve of these sources lie outside our defined CMD cuts and are therefore not included in our initial dataset, but we have manually added them to our list, using the photometry published in that work.  All of these sources are classified as extreme AGB candidates.

\begin{figure}
\begin{center}
\rotatebox{0}{\includegraphics*[scale=0.35]{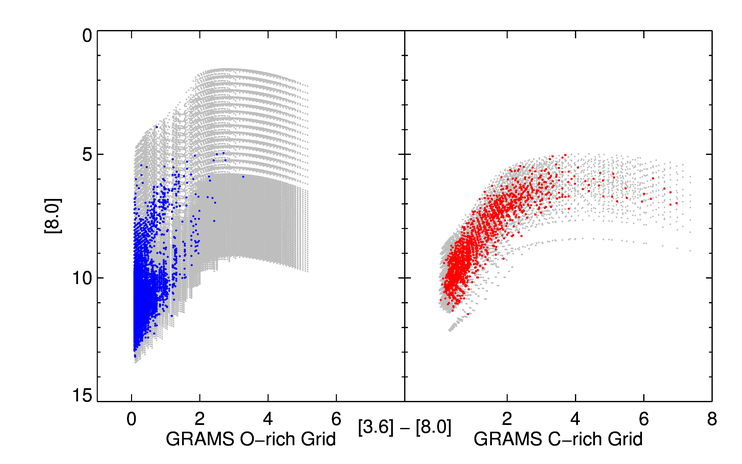}}
\caption{[8.0] vs.\ [3.6] $-$ [8.0] CMD of the entire GRAMS model grid.  \textit{Left panel}: Every model of the O-rich GRAMS grid (light gray) with models selected as a best-fit to one of our sources highlighted in blue.  \textit{Right panel}: Same plot for the C-rich GRAMS grid, with used models highlighted in red.  The two grids were designed to span a larger area of parameter space than real stars are expected to cover.  This plots shows that best-fit models are clustered in both grids and that the grids cover more CMD space than necessary.}
\label{fig:grams_cmd}
\end{center}
\end{figure}

\citetalias{Sargent2011} and \citetalias{Srinivasan2011} present the development of GRAMS, a grid of RT models of dusty evolved stars calculated using the \2dust\ RT code \citep{Ueta2003}.  The dust properties for these models were fixed by modeling two O-rich AGB stars \citep[][hereafter Paper II]{Sargent2010} and one C-rich star in the LMC \citep[][hereafter Paper III]{Srinivasan2010}.  GRAMS consists of 68\,600 O-rich and 12\,243 C-rich models, spanning a large parameter space of stellar photosphere and circumstellar dust shell properties.  The output from GRAMS consists of spectra as well as synthetic photometry for a large set of narrow- and broadband filters, including the ones used in our study. Using these pre-computed SEDs, we are able to find the best fit model to each source in our sample through a simple brute force search on an average desktop computer in only 4 hours.  This approach may reduce the detailed accuracy of any particular model fit, but will compensate by allowing statistically accurate trends and patterns to be determined for entire stellar populations in computationally reasonable periods of time.  Figure~\ref{fig:grams_cmd} shows the entire GRAMS grid on a [8.0] vs.~[3.6] $-$ [8.0] CMD.  The O-rich grid is shown in the left panel and the C-rich grid is shown in the right panel.  The gray background is the entire extent of the grid, and the highlighted points (blue for O-rich, red for C-rich) emphasize the models that have been matched to sources in our dataset.

\subsection{Variability} \label{sec:var}
The overwhelming majority of AGB stars are variable \citep{Vijh2009}, with complicated, multi-periodic light curves \citep[][]{Fraser2008}.  Because our dataset combines multi-epoch observations from different surveys taken on uncorrelated dates, the optical, near-IR and mid-IR portions of the observed SEDs for our stars sample different points in the stars' light curves.  Accurate phase correction for such complicated light curves for a sample the size of ours is not practical, nor possible for the many sources which lack variability data from long-term monitoring surveys like MACHO.   As a major goal of this project was to demonstrate large-scale radiative transfer modeling, and the fact that many of the sources without MACHO data are the extreme AGB candidates which dominate the mass-return to the ISM we seek to measure \citepalias{Srinivasan2009}, we opted to handle the sources' variability in a statistical manner rather than restrict our sample to those objects for which precise phase-correction could be determined.

To account for variability, we took the SAGE mid-IR photometry as the baseline observations, and inflated the error bars on the flux measurement in the shorter wavelength bands (UBVIJHK) by adding an additional error term representative of the source's variability amplitude to the photometric errors.  Because the MACHO survey used a non-standard filter set \citep{Alcock1999}, we first transformed the MACHO b-band photometry into a Johnson V magnitude using the prescription in \citet{Alcock1999}, $$V = 24.114 + 1.00258 b - 0.153 (b-r)$$ where $b$ and $r$ are the MACHO mean b- and r-band magnitudes, respectively.  This transformation allowed us to easily cast the MACHO variability amplitude (originally in magnitude units) as a ratio of the source's flux at minimum brightness to that at maximum brightness.  We choose the MACHO b-band because it is typically larger in amplitude than the r band, and thus we err on the side of caution when modifying our error bars.

We modeled the sources as being a constant flux source with a single period sine-wave signal imposed on it.  We used the b-band mean MACHO magnitude as the average brightness of the source, due to the seven-year baseline of this measurement.  The amplitude of the variation imposed on this mean was the MACHO b-band amplitude associated with the dominant period of variation, taken from \citet{Fraser2008}.
It can be shown that the RMS average of a sine function imposed on a constant flux $F_{V}$ is $$\sigma_{\rm var} = F_{V}\left(\frac{1-\alpha}{1+\alpha}\right)$$ where $\alpha$ is the ratio of the minimum to maximum flux, i.e. the MACHO amplitude (measured in magnitudes) converted to flux units.  One-half this quantity was added in quadrature to the photometric errors of all non-Spitzer bands for each source with available MACHO variability information.  The factor of one-half accounts for the fact that the variability amplitude of AGB stars is greater when measured in the optical (i.e. at the wavelengths studied by MACHO) than in the near-IR \citep[see, eg.][]{Reid2002}.   For those sources in the SAGE catalog without MACHO observations, a ``canonical" variability amplitude was constructed from the sources which did.  All sources were classified as O-rich, C-rich, or extreme AGB candidates, using the photometric criteria described in \S~\ref{sec:dat_source}.  The median amplitude of each class was then assigned to all stars of that class without MACHO data, and we inflated the error bars of the non-SAGE photometry in the same manner discussed above.  That is, the median variability error term of all MACHO-detected sources classed as O-rich AGB candidates was used as the ``canonical" variability error term for all O-rich AGB candidates without MACHO detections, and similarly for C-rich and extreme AGB candidates.  RSGs are not typically as variable as AGB stars, and we do not inflate the errors of RSGs candidates.

\subsection{Fitting Procedure} \label{sec:fitting}
\begin{figure}
\begin{center}
\rotatebox{0}{\includegraphics*[scale=0.35]{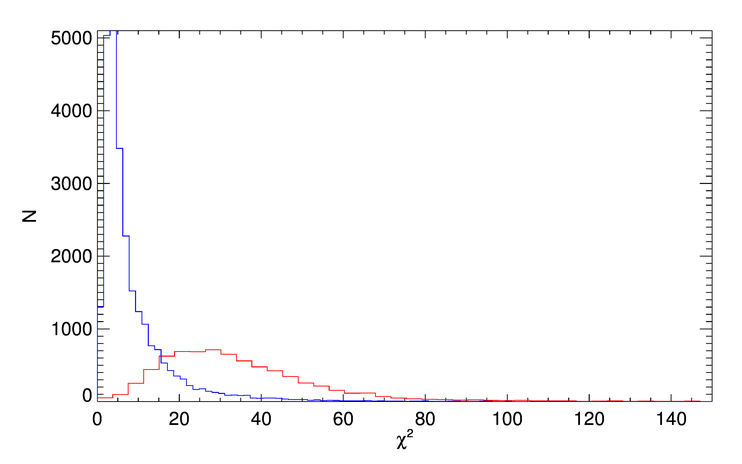}}
\caption{Distribution of the $\chi^2$ per data point for the best-fit model to each source in both the O-rich (blue) and C-rich (red) populations in our sample.  Both distributions are sharply peaked, with long tails extending to large values.  The C-rich sample in general has larger values of $\chi^2$ for the best-fit models.}\label{fig:chi}
\end{center}
\end{figure}

The best-fit GRAMS model for each of our \numsamp\ candidate AGB and RSG stars was found using a brute-force minimum $\chi^2$ search.  Each source was compared to all of the $\sim$68\,000 O-rich GRAMS models, and the $\sim$12\,000 C-rich GRAMS models.  The best-fit model was defined to be the model with the smallest value of the quantity $$\chi^2 = \frac{1}{N}\sum_i \frac{(f_{{\rm obs}_i} - f_{m_i})^2}{\sigma_i^2}$$ where $f_{{\rm obs}_i}$ and $f_{m_i}$ are the observed and model flux in the $i^{\rm th}$ band, respectively.  $N$ is the number of bands for which a source has valid photometry, and $\sigma_{i}$ is the quadrature sum of the photometric extraction error and the variability error term described in \S~\ref{sec:var}.  This quantity is thus properly a $\chi^2$ per data point.  For the reddest sources, defined as those having a J/I flux ratio of 10 or higher (equivalent to $(I-J) > 1.4$), we obtained better results by neglecting their optical (UBVI) photometry entirely, which represents a negligible fraction of the energy of their SEDs.  The J/I flux ratio was selected because this bridges the gap between the MCPS and 2MASS surveys, and is thus sensitive to both sources which are intrinsically very red and variable sources observed by the two surveys at very different points in their light curve.  A star is classified as O- or C-rich based on whether the best fitting model is from the O- or C-rich grid.  RSGs are not a separate classification within GRAMS, but the O-rich grid is designed to cover stars of higher luminosities than the classical AGB limit, including RSGs.  When RSGs are discussed in this work, their classification as such is purely based on the CMD criteria of \citet{Boyer2011} (illustrated in Figure~\ref{fig:cmd_class}).  Figure~\ref{fig:chi} shows the distributions of the $\chi^2$ values for model fits of both types. The median of the O-rich chi-squared distribution is much lower than that of the C-rich one. The differences in these distributions are taken into account when we address the reliability of classifications made by comparing the chi-squared values of the C-rich and O-rich best fit models to a given source (see \S~\ref{sec:oc}).

\begin{figure*}
\begin{center}
\begin{tabular}{cc}
\rotatebox{0}{\includegraphics*[scale=0.25]{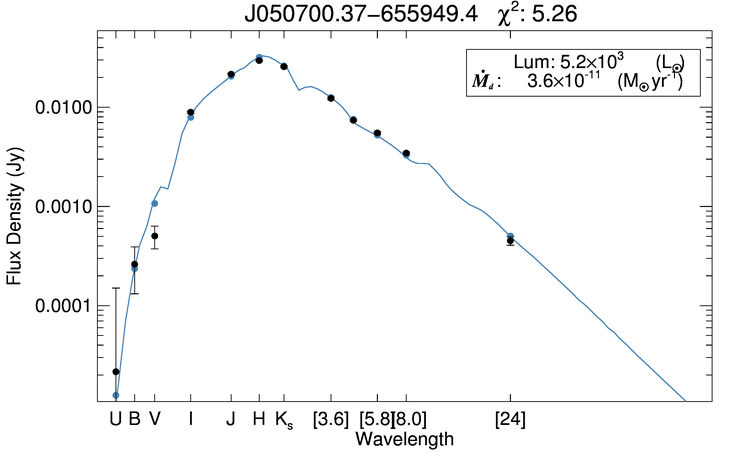}} & \rotatebox{0}{\includegraphics*[scale=0.25]{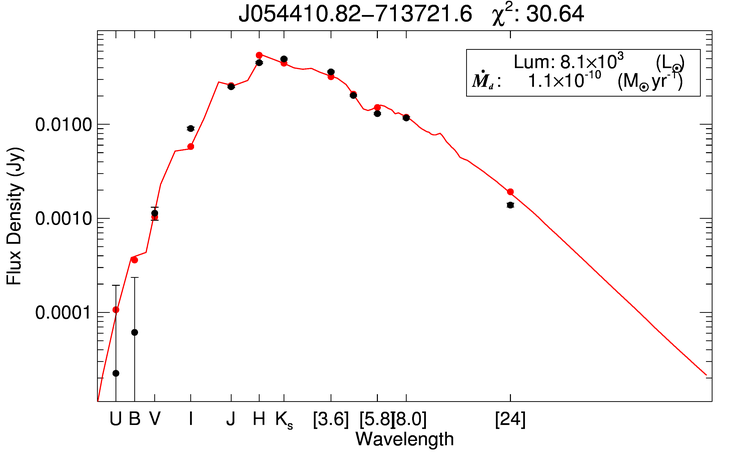}} \\
\rotatebox{0}{\includegraphics*[scale=0.25]{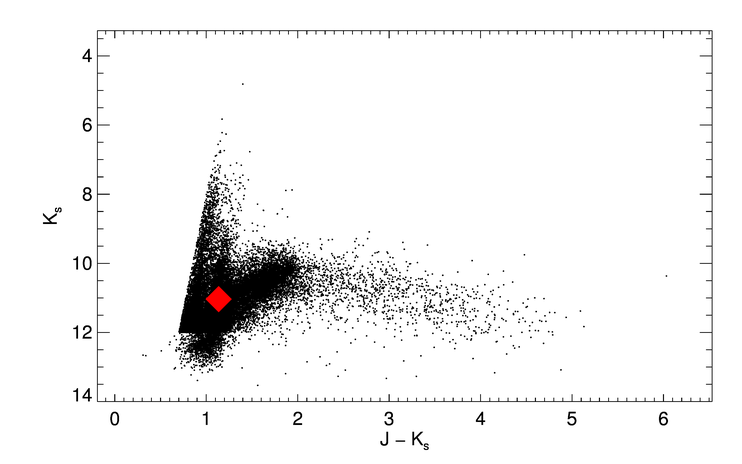}} & \rotatebox{0}{\includegraphics*[scale=0.25]{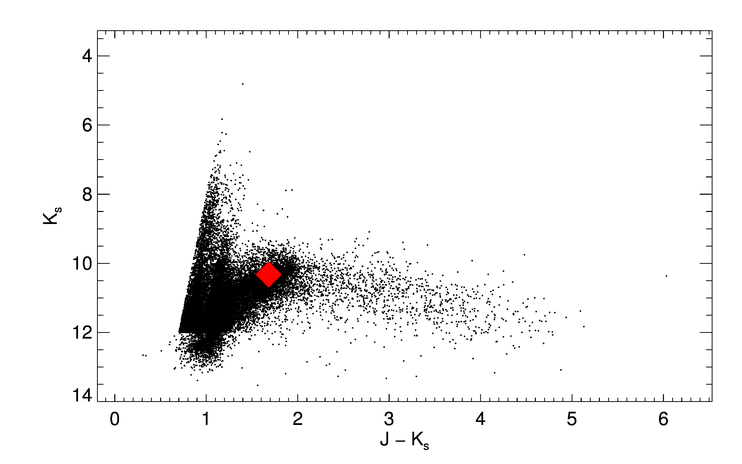}} 
\end{tabular}
\caption{Two example fits from the GRAMS grid.  These particular sources were selected because the quality of their fit ($\chi^2$) more closely matches the median value of $\chi^2$ for their respective grid than any other source.  That is, they are the most `typical' fits in our sample.  \textit{Left panel}: An O-rich source.  \textit{Right Panel}: A C-rich source.  The bottom row shows the location of each source on the \ks\ vs.~\jmag\ $-$ \ks\ CMD. }
\label{fig:ave_fit}
\end{center}
\end{figure*}

Figure~\ref{fig:ave_fit} shows two typical fits to our data.  These fits are `typical' in that they were specifically selected to have values of $\chi^2$ closest to the median value for their class (O- or C-rich).  The left column shows the SED of the median O-rich fit above its location in the \ks\ vs.~\jmag\ $-$ \ks\ CMD.  The right column shows the same plots for the prototypical C-rich fit.  Both of these sources are centrally located in their respective populations in the IR CMD (see Figure~\ref{fig:cmd_class}).  The effects of inflating the non-SAGE photometric error bars to account for variability can clearly be seen in the U- and B-band fluxes in the C-rich SED, right column.

After this initial fitting, we visually inspected the SEDs of several hundred sources.  The distribution of $\chi^2$ for both the O-rich and the C-rich samples were found to be sharply peaked (Figure~\ref{fig:chi}), and we selected the 5\% of the sample with the largest value of $\chi^2$ for inspection.  For these stars we visually inspected the fit derived using all valid photometry and the fit to the IR (2MASS and SAGE) data only.  If the IR fit was significantly better (both quantitatively as defined by the value of $\chi^2$ and qualitatively as defined by eye) the source was manually flagged to be fit using only the IR bands.  Based on both manual identification and color criteria (J/I flux ratio), \numir\ of our sources were fit using IR data only.  We flagged \numbad\ sources as invalid fits.  

\begin{figure*}
\begin{center}
\begin{tabular}{cc}
\rotatebox{0}{\includegraphics*[scale=0.25]{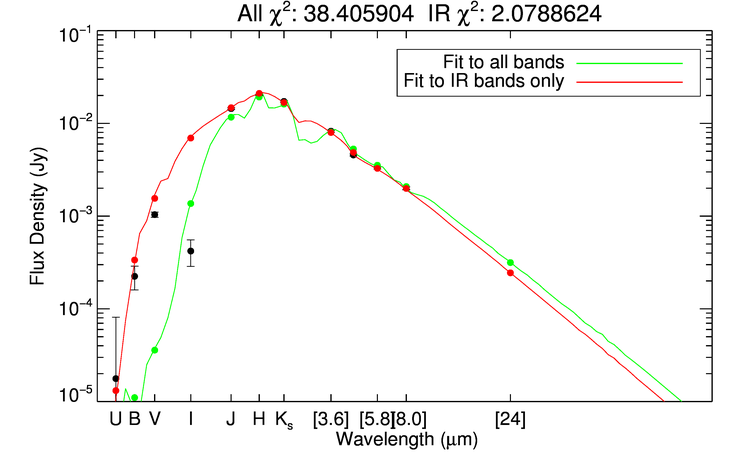}} & \rotatebox{0}{\includegraphics[scale=0.25]{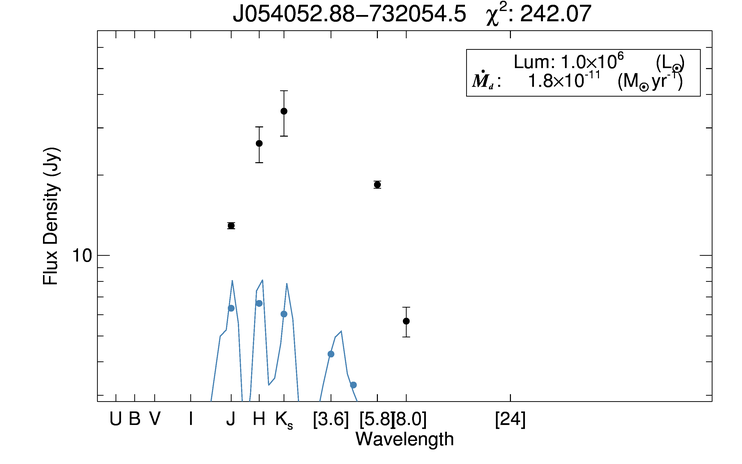}} \\
\rotatebox{0}{\includegraphics*[scale=0.25]{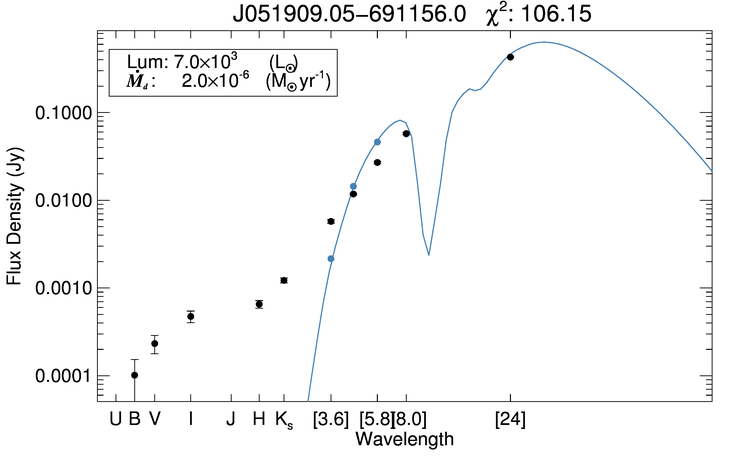}} & \rotatebox{0}{\includegraphics[scale=0.25]{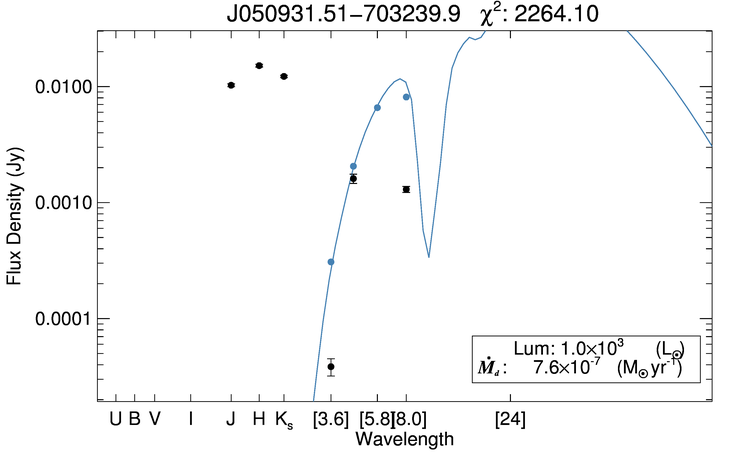}} 
\end{tabular}
\caption{Four sources selected for visual inspection of the fit.  \textit{Top left}: A source with an anomalously dim I-band flux. 202 sources had similar SEDs.  The actual photometry is plotted in black, the best-fit model based on all valid photometry is in green, and the fit to only the IR data (2MASS and Spitzer) is in red.  Notice the IR-based fit does a better job of matching all the available photometry because it is not skewed by the I-band.  This source was kept in our sample, using the fit to only the IR data.  \textit{Top Right}: The unphysical brightness of this source indicates it is not at the distance of the LMC.  This source was removed from our dataset.  \textit{Bottom Left}:  A source with an SED inconsistent with an AGB star.  The monotonic increase of the flux of this star as one moves to longer wavelengths is not typical of AGB stars, which peak in the near-IR.  This source was removed from our dataset.  \textit{Bottom Right}:  A source with bad data quality.  The bizarre SED of this star is most likely due to a mismatch between the optical and IR surveys used in this study.  This source was removed from our dataset.}
\label{fig:bad_fit}
\end{center}
\end{figure*}

The discarded sources included those with too few valid data points ($\leq 4$ bands), obvious foreground sources, and sources with SEDs not consistent with an AGB star.  Figure~\ref{fig:bad_fit} displays some example sources with fits that were manually adjusted or rejected.  At top left we show a source with an anomalously dim $I$-band flux.  There were 202 such sources in our dataset.  Because the fit based only on the IR data points (red curve) matches the overall SED much better than that based on all the photometry including the $I$-band (green curve), these sources were retained in our dataset, using the fit based only on the IR data.  At the top right of Figure~\ref{fig:bad_fit}, we have a source which is obviously a foreground source, mis-identified in our catalog as belonging to the LMC.  The best fitting model (blue curve) has fluxes $\sim$10 times dimmer than the observed object, yet a luminosity of a million~L$_{\odot}$.  If this source were truly in the LMC, it would be unphysically luminous.  At the bottom right, we see a source with poor data quality.  Such bright 2MASS photometry compared to such dim [3.6] and [4.5]~\mic\ photometry is probably the result of a mismatch in the join between the 2MASS and SAGE catalogs.  Such sources were removed from our sample.  Finally, at the bottom left of Figure~\ref{fig:bad_fit}, we see a source with an SED not typical of an AGB star or RSG.  The anomalous optical photometry indicates that this source is either a mismatch between catalogs, or not an evolved star.

We are left with a dataset of \numvalid\ sources matched to GRAMS models from which we obtain astrophysical parameters such as dust mass-loss rate (\mlr), luminosity, and effective temperature (\teff).  Table~\ref{tab:data} summarizes these numbers.  Our entire \numvalid\ source catalog is available online as a machine readable table.  Table~\ref{tab:source_list} presents a small excerpt as a guide to the structure of the electronic table.  For each source, we list the SAGE mosaic photometry ID, the classification assigned to the source using the CMD criteria of \citet{Cioni2006}, the classification assigned by GRAMS, the $\chi^2$ of the best-fitting model, the \mlr, the luminosity, the \teff\, and the optical depth ($\tau$) of the circumstellar dust shell (determined at 10~\mic\ if a star is classified O-rich, 11.3~\mic\ if C-rich) of that model, along with the associated uncertainties.  MACHO variability data (amplitude and period) are included for sources for which it is available, along with the UBVIJHK, [3.6], [4.5], [5.8], [8.0], and [24]~\mic\ fluxes (and uncertainties) for every source.  The UBVIJHK photometry listed in the online table are not dereddened.  Photometric uncertainties have not been inflated as described in \S~\ref{sec:var}.

\begin{deluxetable}{lcc}
\tabletypesize{\scriptsize}
\tablecaption{Dataset Populations}
\tablecomments{Numbers of sources classified as each type (O-rich, C-rich, Extreme AGB, and RSG) using both color-magnitude cuts \citep[see][for details]{Riebel2010} and using the GRAMS model grid.  GRAMS classifications are based on the class of the single best-fitting model, as defined by the smallest value of $\chi^2$.}
\tablehead{ \colhead{} & \colhead{Class (CMD)} & \colhead{Class (GRAMS)} }
\startdata
O-rich\tablenotemark{a} &        19\,566 &        26\,210 \\
C-rich AGB 				&         6709   &         7281 \\
Extreme AGB 			&         1340   & 			N/A \\
Red Supergiant 			&         5876   & 			N/A 
\enddata
\tablenotetext{a}{The GRAMS O-rich classification includes RSGs, as well as AGB stars}
\label{tab:data}
\end{deluxetable}

\begin{deluxetable*}{lcccccccccccccc}
\tabletypesize{\scriptsize}
\tablecolumns{17}
\tablewidth{0pt}
\tablecaption{Source List}
\tablecomments{Source list of dataset used in this study.  The full dataset is available in the online materials of this paper.  This excerpt is included only as a guide to format.}
\tablehead{ \colhead{SAGE ID\tablenotemark{a}} & \colhead{Class\tablenotemark{b}}  & \colhead{Class\tablenotemark{c}} & \colhead{$\chi^2$} & \colhead{\mlr} & \colhead{$\sigma_{\mbox{\mlr}}$} & \colhead{L$_{\rm bol}$} & \colhead{$\sigma_{L}$} &   \colhead{Amplitude} & \colhead{Period} &  \colhead{\ldots \tablenotemark{d}} \\
 \colhead{} &  \colhead{(CMD)} & \colhead{(GRAMS)} &\colhead{} & \colhead{(\msunperyr)} & \colhead{(\msunperyr)} & \colhead{(L$_{\odot}$)} & \colhead{(L$_{\odot}$)} &  \colhead{(mag)} & \colhead{(days)} & }
\startdata
J050115.85-692040.4 & o & o &   5.56 &  5.71E-11 &  1.19E-11 &    4149 &  299 &    0.13 & 783.70 &   \ldots \\
J050311.47-691412.3 & o & c &  17.36 &  4.22E-11 &  1.69E-11 &   10000 &  879 &    0.08 & 110.61 &   \ldots \\
J051227.54-701730.7 & x & c &  42.53 &  4.22E-10 &  5.69E-11 &    5128 &  384 &    1.74 & 356.63 &   \ldots \\
J060647.79-664812.5 & s & o &  31.52 &  1.33E-08 &  9.16E-09 & 1000000 &  -99 &    0.00 &   0.00 &   \ldots 
\enddata
\tablenotetext{a}{In the online table, all SAGE-IDs are prefaced with `SSTISAGEMA'}
\tablenotetext{b}{CMD Classifications are described in \citet{Cioni2006} and \citet{Riebel2010}}
\tablenotetext{c}{O-rich or C-rich classification based on best fitting GRAMS model}
\tablenotetext{d}{The online data also contain the optical depth, T$_{\rm eff}$, UBVIJH\ks, [3.6], [4.5], [5.8], [8.0], [24] fluxes, and their associated uncertainties}
\label{tab:source_list}
\end{deluxetable*}

\begin{figure}
\begin{center}
\rotatebox{0}{\includegraphics*[scale=0.35]{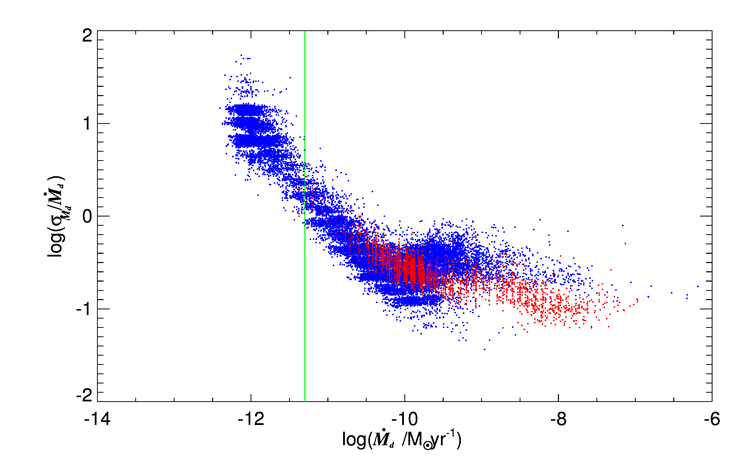}}
\caption{$\log(\sigma_{\dot{M_{d}}}/\dot{M_{d}})$ vs. $\log(\dot{M_{d}})$ for our sample.  O-rich sources in blue, C-rich in red.  O-rich sources to the left of the green line at $\log(\dot{M_{d}})=-11.3$ are classified as low-\mlr\ sources (\S~\ref{sec:mlr}) and are consistent with $\dot{M_{d}}=0$.  All of these low-\mlr\ sources have $\sigma_{\dot{M_{d}}}/\dot{M_{d}} > 1$ while only 6\% of sources with higher values of \mlr\ do.}
\label{fig:sigma_mlr}
\end{center}
\end{figure}

The single best-fit model to a given source does not have uncertainty in its parameters such as luminosity, \mlr, and \teff\ (strictly speaking, \mlr\ is not a parameter of the model, but it is an immediate consequence of optical depth, which is).  In order to define an uncertainty for these parameters, we use the median absolute deviation (MAD) for each parameter from the 100 best fitting models in the same grid (O-rich or C-rich) as the uncertainty for that parameter.  That is, for any given model parameter $X$ (\mlr, luminosity, etc.), the quantity $\sigma_{X} \equiv {\rm Median}(|X_{i} - {\rm Median}(X_{i})|)$, where \textit{i} ranges over the 100 best-fitting models, is calculated and defined as the uncertainty in that parameter.  The MAD is a more robust estimate for the spread in parameter values than the standard deviation or a similar statistic, and more appropriate for our purposes since it is highly unlikely that all of the output parameters from the best-fitting 100 models to a given source will be normally distributed.  Note that in the case of normally distributed errors, the MAD is smaller than the standard deviation, specifically $\rm{std dev} \approx 1.5({\rm MAD})$.  For some of the more coarsely sampled parameters, such as \teff, all 100 best-fitting models may have identical values of a given parameter.  In these cases, the MAD will be 0, and in Table~\ref{tab:source_list} we have set the uncertainty to the unambiguous placeholder value $-99$.  The number 100 was settled on after experimentation with various values from 10--1000.  The number 100 is on the order of 1\% of the GRAMS grid for both C-rich and O-rich models, and ensures that the majority of sources in our sample will have well defined uncertainties (not $-99$) for most parameters.  Smaller subsets do not sample a large enough region of parameter space, and often have all models in the subsample with identical values for many parameters.  Larger subsets sample poorly fitting models and can cause an unacceptable number of models to have $\sigma_{X}/X >1$.  Using 100 models to define the uncertainty is a compromise between these two positions, with emphasis placed on the uncertainty in \mlr. The number of sources with valid errors for each parameter are listed in Table~\ref{tab:err_num}.  We see that most sources have well-defined uncertainties for \mlr\ and luminosity.  The uncertainty in \teff\ is not well-determined from our model grid.  This is an expected result.  As discussed in \citetalias{Sargent2011} and \citetalias{Srinivasan2011}, the model photospheres used in generating the GRAMS grid do not offer a very fine sampling of effective temperatures.    Figure~\ref{fig:sigma_mlr} shows the logarithm of the ratio of the uncertainty in \mlr\ to the \mlr\ itself, $\log(\frac{\sigma_{\dot{M_{d}}}}{ \dot{M_{d}}})$ vs.\ $\log(\dot{M_{d}})$.  O-rich sources with $\log(\dot{M_{d}}) < -11.3$ (to the left of the green line in Figure~\ref{fig:sigma_mlr}) are not thought to represent actual physical circumstellar dust shells (see \S~\ref{sec:mlr} for discussion).  Significantly, this value of \mlr\ is also approximately where the S/N equals unity.  Only 6\% of sources with $\log (\dot{M_{d}}) \ge -11.3$ have a signal-to-noise less than 1, while 100\% of the low-\mlr\ sources do.

\begin{deluxetable*}{lrrrr}
\tabletypesize{\scriptsize}
\tablewidth{0pt}
\tablecaption{Sources with Valid Uncertainties}
\tablecomments{Number of sources with valid uncertainties for each parameter obtained from the model grid.  Uncertainties in model parameters are defined as the median absolute deviation of that parameter for the 100 best-fitting models to a given source.  If all 100 models have the same value for a given parameter, the median absolute deviation is mathematically $0$ and a placeholder value of $-99$ is listed.}
\tablehead{ \colhead{Class (GRAMS)} & \colhead{Luminosity} & \colhead{\mlr} & \colhead{T$_{\rm eff}$} & \colhead{$\tau$} }
\startdata
O-rich &        25\,579 (97.6\%) &        26\,210 (100\%) &        7710 (70.6\%) &        25\,626 (97.8\%) \\
C-rich &         6963 (95.6\%) &         7281 (100\%) &        6996 (96.1\%) &         6815 (93.6\%) 
\enddata
\label{tab:err_num}
\end{deluxetable*}

\section{Results: SED Fits and Types of Sources} \label{sec:tour}
We fit GRAMS models to the SEDs of the O-rich stars (SSTISAGE1C J052206.92-715017.6 and HV 5715), as well as the C-rich star OGLE LMC LPV 28579. These sources were modeled in detail in \citetalias{Sargent2010} and \citetalias{Srinivasan2010} in order to establish the dust properties for the grid. Our results, listed in Table~\ref{tab:ben_sundar}, agree to within uncertainties on almost every model parameter for all three sources.  The exception, optical depth, is due to coarseness in the model grid coverage.  The fact that our population scale fitting procedures can match the output of detailed modeling so well gives us a great amount of confidence in the overall accuracy of our results.

\begin{deluxetable*}{lcccccc}
\tabletypesize{\scriptsize}
\tablewidth{0pt}
\tablecaption{Comparison of fitting results for sources also in \citetalias{Sargent2010} and \citetalias{Srinivasan2010}}
\tablecomments{Comparison of the stellar parameters for the sources examined in \citetalias{Sargent2010} and \citetalias{Srinivasan2010}.  Each source is listed twice, first with the parameter values found in this work, and second with the values found in the previous study in which it was modeled more precisely.  If no uncertainty was found for a parameter (see \S~\ref{sec:fitting}), none is listed.  Where uncertainties are not symmetric about the best fit value, the range is given in parentheses.}
\tablehead{ \colhead{SAGE ID\tablenotemark{a}} & \colhead{ID\tablenotemark{b}} & \colhead{GRAMS Class} & \colhead{\mlr\ ($\times 10^{-9}$\msunperyr)} & \colhead{Luminosity (L$_{\odot}$)} & \colhead{\teff (K)} & \colhead{$\tau$\tablenotemark{c}} }
\startdata
J051811.08-672648.5\tablenotemark{d} & 82 & O &$1.5\pm 0.4$   & $33\,694 \pm  6000$ & $3500 \pm 400$ & $0.0256$ \\
 \hspace{0.5cm}(HV 5715)             &    & O &$2.3(1.1-4.1)$ & $36\,000 \pm 4000$  & $3500 \pm 100$ & $0.012 \pm 0.003$ \\
J052206.92-715017.7\tablenotemark{e} & 96 & O &$2.1\pm0.4$    & $4820 \pm 670$      & $3700 \pm 200$ & $0.1024$ \\
 \hspace{0.5cm}(SSTSAGE052206)       &    & O &$2.0(1.1-3.1)$ & $5100 \pm 500$      & $3700 \pm 100$ & $0.095(0.07-0.13)$ \\
J051306.40-690946.3\tablenotemark{f} & 66 & C & $2.1\pm 0.4$  & $7080 \pm 700$      & $3100 \pm 200$ & $0.4$ \\
 \hspace{0.5cm}(LPV 28579)           &    & C & $2.5(2.5-2.9)$ & $6580(6150-7010)$   & $3000$         & $0.27(0.25-0.275)$ 
\enddata
\tablenotetext{a}{Source IDs from this work are all prefaced with `SSTISAGEMA'}
\tablenotetext{b}{ID from the SAGE-Spec survey}
\tablenotetext{c}{The quoted optical depth is that at 10.0~\mic\ for sources in \citetalias{Sargent2010} and at 11.3~\mic\ for the source from \citetalias{Srinivasan2010}.}
\tablenotetext{d}{Source HV 5715 from \citetalias{Sargent2010}}
\tablenotetext{e}{Source SSTSAGE052206 from \citetalias{Sargent2010}}
\tablenotetext{f}{Source LPV 28579 from \citetalias{Srinivasan2010}}
\label{tab:ben_sundar}
\end{deluxetable*}

\begin{figure*}
\begin{center}
\begin{tabular}{cc}
\rotatebox{0}{\includegraphics*[scale=0.25]{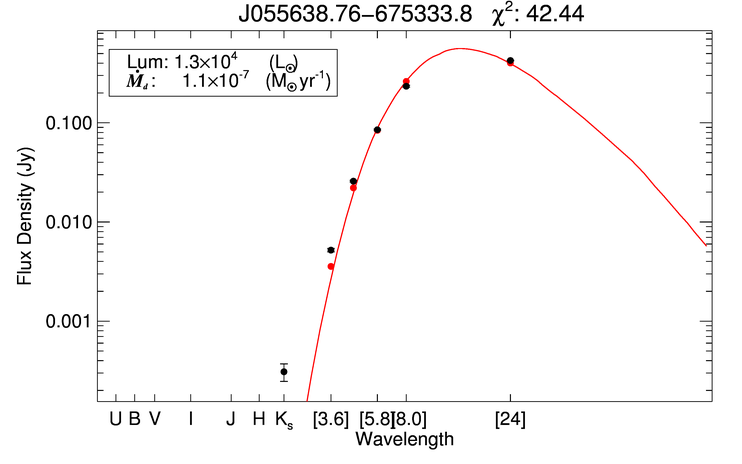}} & \rotatebox{0}{\includegraphics[scale=0.25]{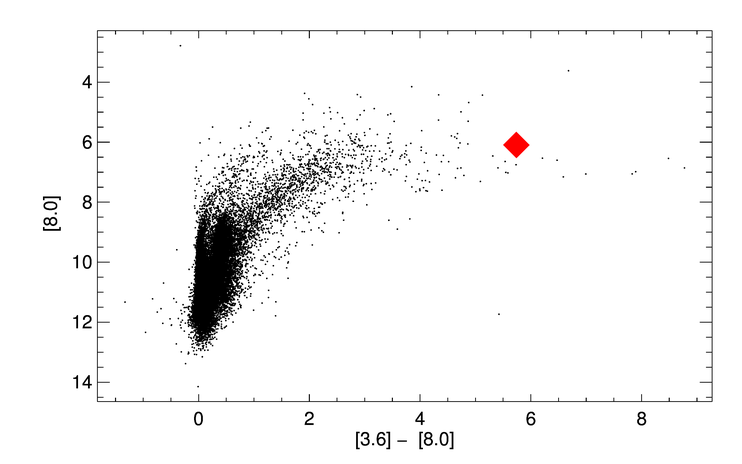}} \\
\rotatebox{0}{\includegraphics*[scale=0.25]{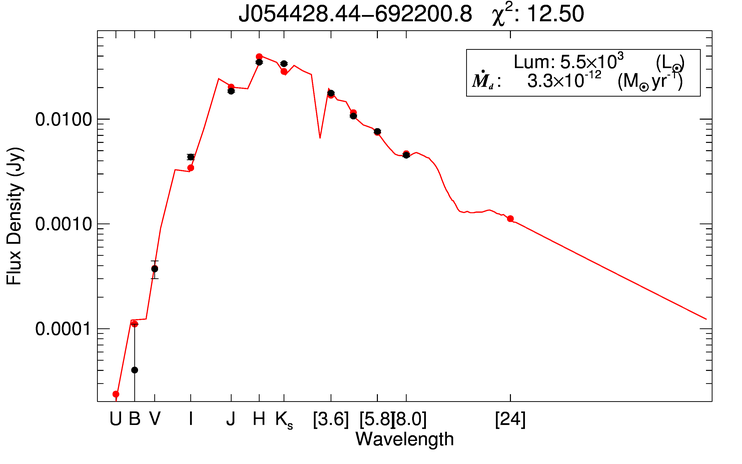}} & \rotatebox{0}{\includegraphics[scale=0.25]{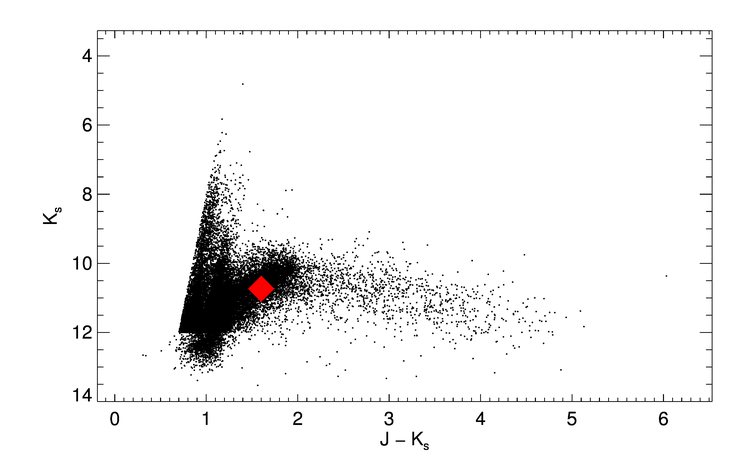}} \\
\rotatebox{0}{\includegraphics*[scale=0.25]{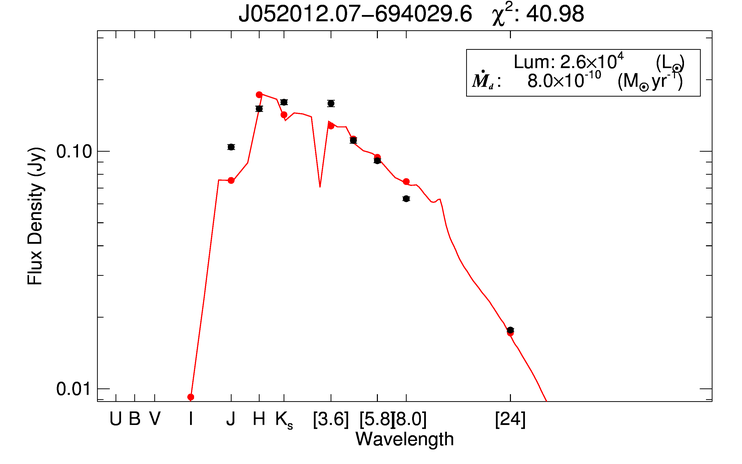}} & \rotatebox{0}{\includegraphics[scale=0.25]{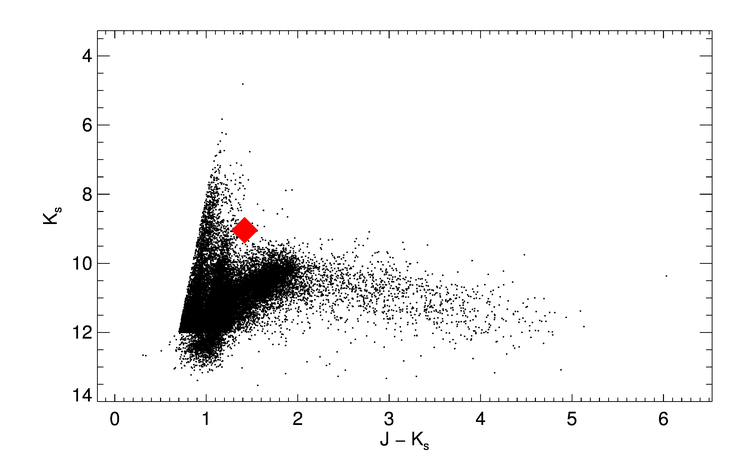}} \\
\rotatebox{0}{\includegraphics*[scale=0.25]{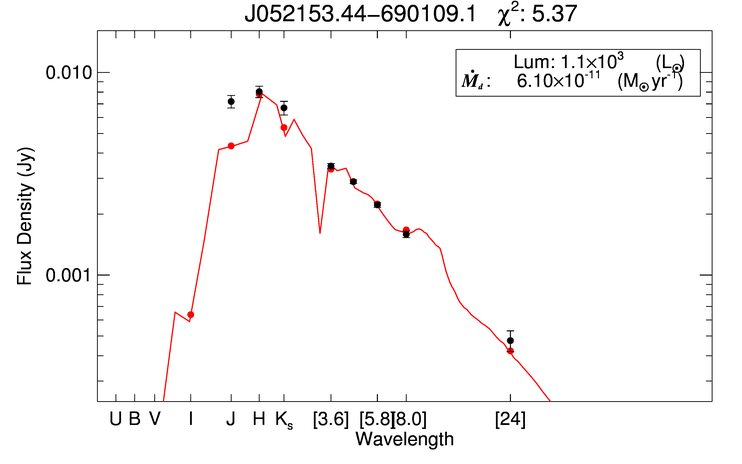}} & \rotatebox{0}{\includegraphics[scale=0.25]{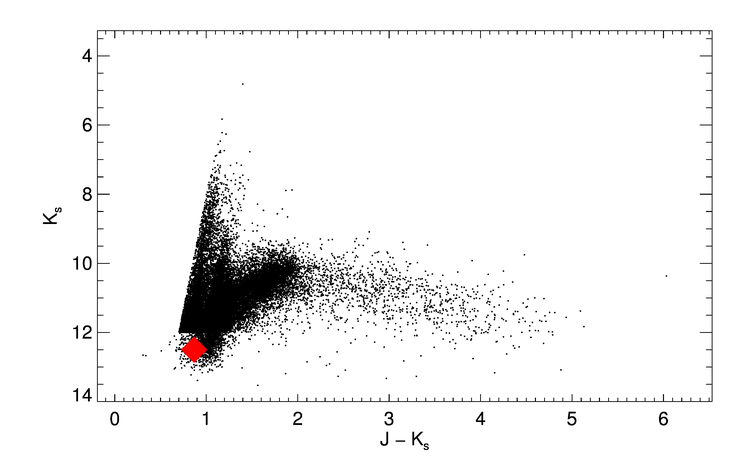}} \\
\end{tabular}
\caption{C-rich sources showing the largest/smallest values of the \mlr\ and luminosity parameters found in our sample.  The left column shows the SED fit for each source, and the right column shows its location in the \ks\ vs.\ \jmag\ $-$ \ks\ CMD (if the 2MASS photometry for a source is not available, the [8.0] vs.\ [3.8] $-$ [8.0] CMD is shown instead).  \textit{Top row}: Highest value of \mlr.  \textit{Second Row}: Smallest values of \mlr.  \textit{Third Row}:  Highest luminosity.  \textit{Fourth Row}: Lowest luminosity.}
\label{fig:span_param_c}
\end{center}
\end{figure*}

\begin{figure*}
\begin{center}
\begin{tabular}{cc}
\rotatebox{0}{\includegraphics*[scale=0.25]{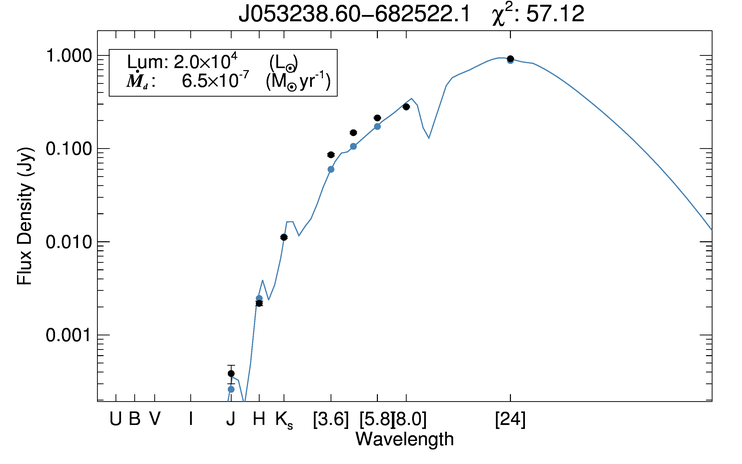}} & \rotatebox{0}{\includegraphics[scale=0.25]{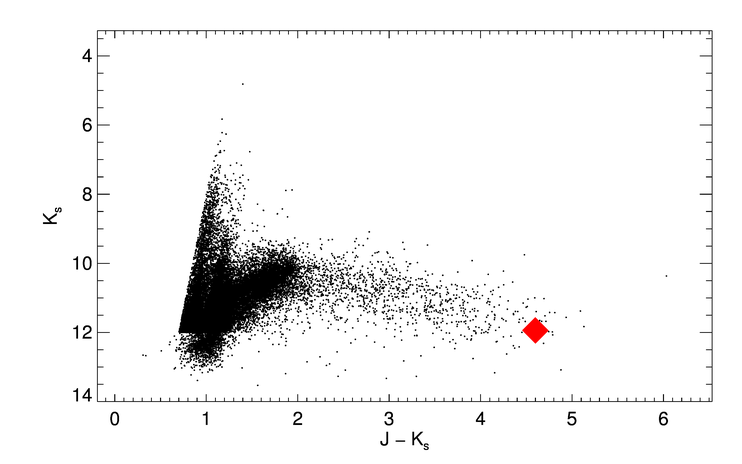}} \\
\rotatebox{0}{\includegraphics*[scale=0.25]{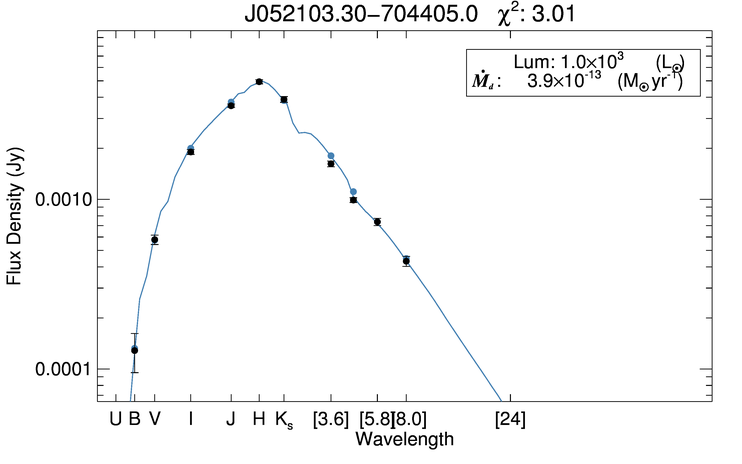}} & \rotatebox{0}{\includegraphics[scale=0.25]{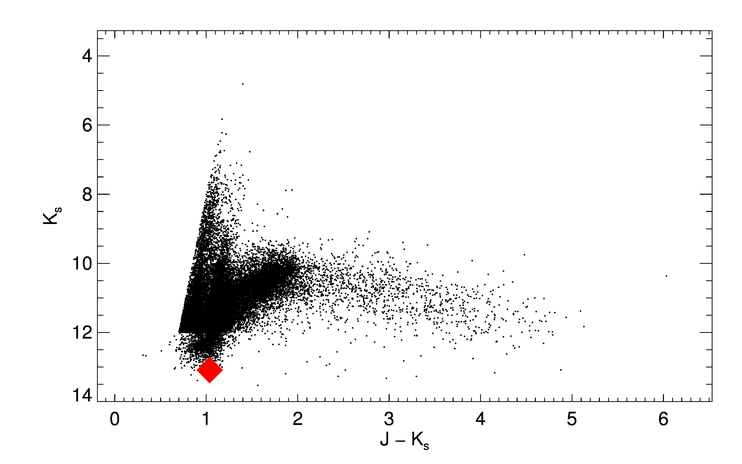}} \\
\rotatebox{0}{\includegraphics*[scale=0.25]{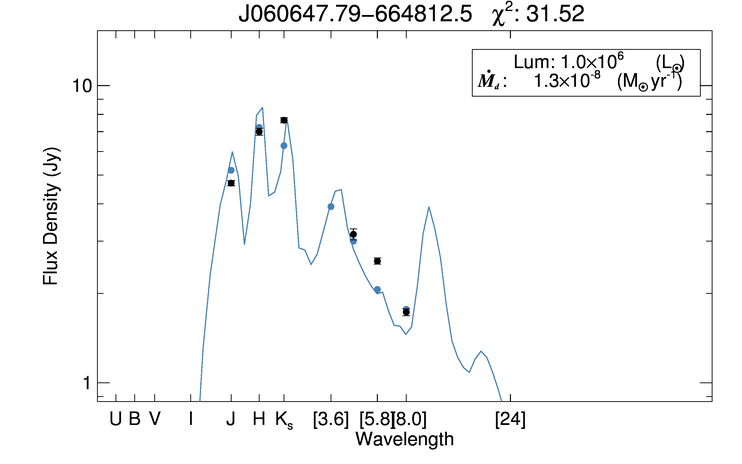}} & \rotatebox{0}{\includegraphics[scale=0.25]{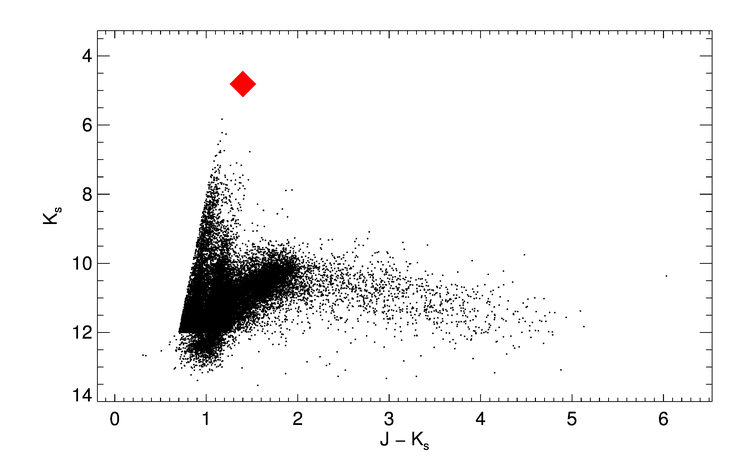}} \\
\rotatebox{0}{\includegraphics*[scale=0.25]{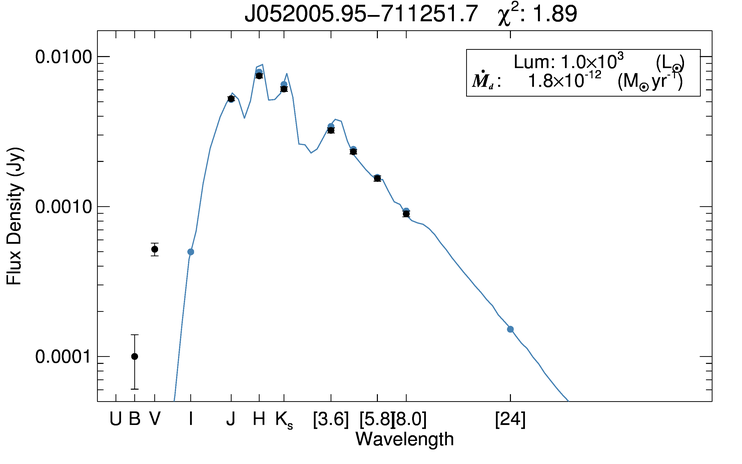}} & \rotatebox{0}{\includegraphics[scale=0.25]{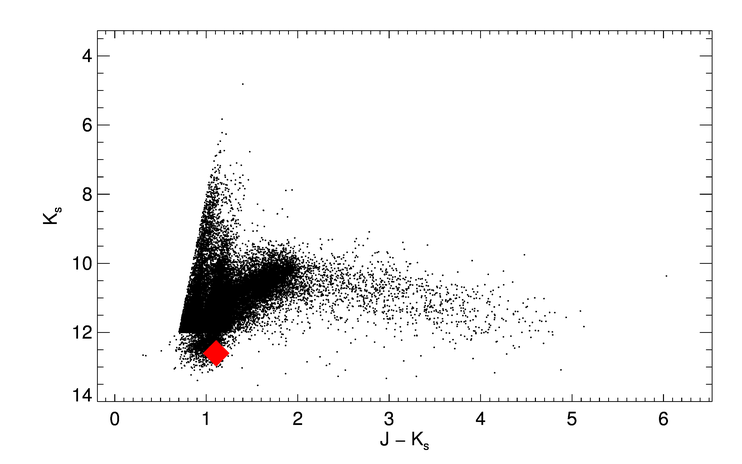}} \\
\end{tabular}
\caption{O-rich sources showing the largest/smallest values of the \mlr\ and luminosity parameters found in our sample.  The left column shows the SED fit for each source, and the right column shows its location in the \ks\ vs.\ \jmag\ $-$ \ks\ CMD (if the 2MASS photometry for a source is not available, the [8.0] vs.\ [3.8] $-$ [8.0] CMD is shown instead).  \textit{Top row}: Highest value of \mlr.  \textit{Second Row}: Smallest values of \mlr.  \textit{Third Row}:  Highest luminosity.  \textit{Fourth Row}: Lowest luminosity.}
\label{fig:span_param_o}
\end{center}
\end{figure*}

Figures~\ref{fig:span_param_c} and \ref{fig:span_param_o} show the SEDs and model fits to 8 sources selected to illustrate the range of stellar parameters spanned by our sample.  Specifically, we show sources with the maximum and minimum values of both \mlr\ and luminosity for each class of source, C-rich and O-rich.  In both figures, the right hand column shows an IR CMD highlighting the specific source shown in the left column. 

The top source in figure~\ref{fig:span_param_c} was chosen because it has the largest value for \mlr\ in our C-rich sample.  Its SED peak is shifted far into the IR, peaking redward of 8~\mic.  It is a bright source, but only half as bright as the brightest C-rich sources in our sample (see below).  This is consistent with Fig~\ref{fig:mlr_lum} (discussed in \S~\ref{sec:mlr}), where there is no clear relationship between \mlr\ and luminosity seen in our sample for either C-rich or O-rich sources.  This source is highlighted in the [8.0] vs.\ [3.6] $-$ [8.0] CMD in the right panel.  This \textit{Spitzer} CMD is used because this source is so red that it is lacking a valid \jmag\ or \ks\ detection, and therefore cannot be shown on the \jmag\ $-$ \ks\ vs. \ks\ CMD.  As one would expect for such a red source with a high \mlr, this source is classified as an extreme AGB star using the color criteria of \citet{Blum2006}.  As discussed in \S~\ref{sec:mlr}, these extreme AGB stars dominate the mass return to the ISM in the LMC, accounting for 74\% of the total amount of mass lost, but only 4\% of the sample by number.  

The source shown in the second row of Figure~\ref{fig:span_param_c} has the lowest \mlr\ of our sample.  It is one of the few stars in our sample classified as C-rich by GRAMS that qualify as a low-\mlr\ star, defined as having $\log(\dot{M_d}) < -11.3$ (\S~\ref{sec:mlr}).  The uncertainty in its \mlr\ ($\pm 1.3\times 10^{-11}$\msunperyr) makes this source consistent with a bare photosphere and zero \mlr.  The SED of this source peaks in the \hmag\ band, consistent with a much hotter object ($\sim$2500~K) such as an actual visible star not surrounded by dust.

The third and fourth rows of Figure~\ref{fig:span_param_c} show sources with the maximum and minimum values (respectively) of luminosity for the C-rich grid.  Both of these sources have similar near-IR colors and similar low values of \mlr.

Fig~\ref{fig:span_param_o} has the same format as Figure~\ref{fig:span_param_c}, but all the sources here are classified as O-rich.  Again, the source's location on an IR CMD is shown in the right column.

The top source in Figure~\ref{fig:span_param_o} has the highest \mlr\ of any O-rich source in our sample.  Its \jmag\ $-$ \ks\ color is extremely red because the SED peak has been shifted all the way to $\sim$24~\mic\ by thick circumstellar dust.  We see a sharp contrast with the low-\mlr\ source (essentially a bare photosphere) seen in the second row.  For the low-\mlr\ source, the SED peaks in the \hmag\ band.

The bottom two rows of Figure~\ref{fig:span_param_o} are the highest and lowest luminosity sources in our O-rich (which includes models of RSGs) sample.  The lowest luminosity source shown at the bottom has a very low \mlr\ as well, but the brightest source ($1\times 10^{6}$~L$_\odot$) shows quite a high rate of luminosity driven mass-loss.

\section{DISCUSSION}\label{sec:discuss}

\subsection{O-Rich/C-Rich Discrimination}\label{sec:oc}
\begin{figure*}
\begin{center}
\begin{tabular}{cc}
\rotatebox{0}{\includegraphics*[scale=0.25]{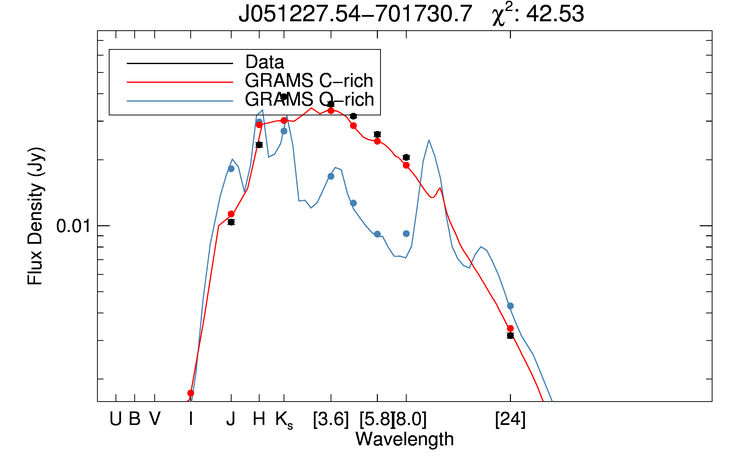}} &\rotatebox{0} {\includegraphics*[scale=0.25]{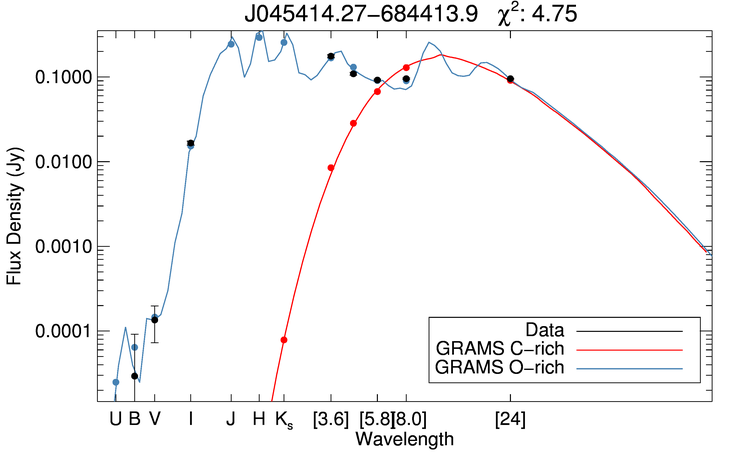}}\\
\rotatebox{0}{\includegraphics*[scale=0.25]{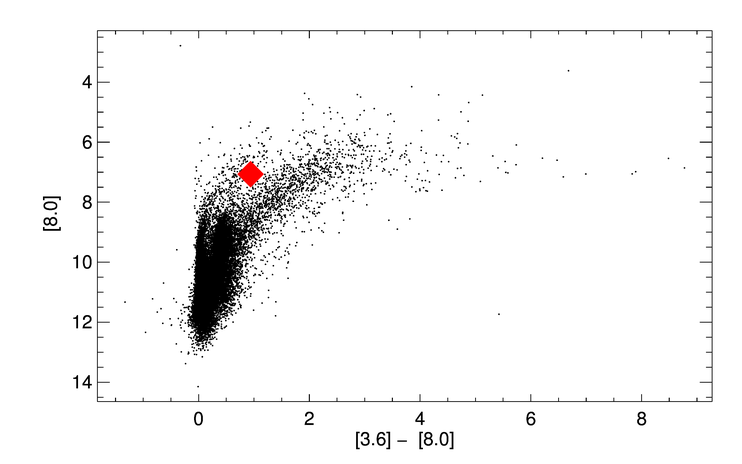}} & \rotatebox{0} {\includegraphics*[scale=0.25]{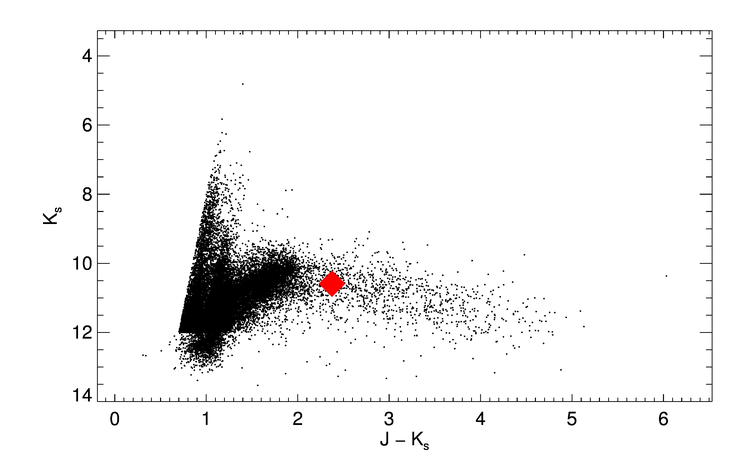}}
\end{tabular}
\caption{Two sources classified as `extreme' AGB stars using the CMD criteria from \citet{Cioni2006}.  In both panels, the blue curve shows the best-fitting O-rich model, and the red curve shows the best-fitting C-rich model.  \textit{Left Panel}: A source classified as C-rich by the GRAMS model grid.  97\% of the extreme sources are classified thus.  \textit{Right Panel}: A source classified as O-rich by the GRAMS model grid.  The bottom row shows the location of the C-rich source in the \ks\ vs.~\jmag\ $-$ \ks\ CMD, and the O-rich source in the [8.0] vs.~[3.6] $-$ [8.0] CMD due to the absence of JHK photometry.}
\label{fig:xtreme_fit}
\end{center}
\end{figure*}

\citet{Cioni2006} proposed a photometric means of discriminating between O- and C-rich AGB stars in the \ks\ vs. \jmag\ $-$ \ks\ CMD.  The GRAMS grid shows good agreement with these color-magnitude cuts, giving us confidence in the grid's ability to reproduce previously established population-scale classification schema.  Table~\ref{tab:data} summarizes our agreement with the 
cuts from \citet{Cioni2006}.  Of the \numo\ sources we classify as O-rich, 95\% are classified as either O-rich AGBs using the cuts from \citet{Cioni2006} or as RSGs using the cuts from \citet{Boyer2011}.  We also obtain a 83\% agreement between sources classified as C-rich by the color-magnitude cuts of \citet{Cioni2006} and sources we match to C-rich models.  The extreme AGB star classification does not exist as a separate category in the GRAMS model grid.  One of the aims of this project was to provide O- or C-rich classification for these sources, most of which are expected to be highly evolved and thus C-rich.  Of the \numx\ sources classified as extreme based on CMD definitions, \numxc\ sources (97\%) are matched to a C-rich GRAMS model.  Figure~\ref{fig:xtreme_fit} shows the SEDs of two sources, both classified as extreme AGB stars by CMD criteria.  The left column shows a star which is best fit by a C-rich model, primarily on the strength of its SAGE photometry, and the right column shows a star classified as an O-rich source, primarily because of its optical photometry.

We find 29 sources (not classified as RSGs via CMD cuts) in our sample with luminosities greater than the classical AGB limit, 54\,000 L$_{\odot}$.  This luminosity limit is based on the point at which the core mass of an AGB star would reach the Chandrasekhar limit, based on the luminosity-core mass relation of \citet{Paczyski1971}.  However, the most massive AGB stars can violate this limit due to hot-bottom burning \citep{Bloecker1991}, which also tends to prevent a star from becoming C-rich \citep{Boothroyd1993}.  

\begin{figure}
\begin{center}
\begin{tabular}{c}
\rotatebox{0}{\includegraphics*[scale=0.35]{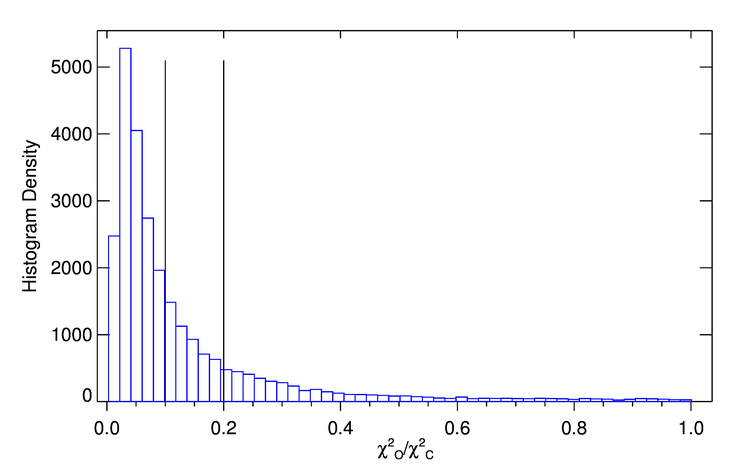}} \\
\rotatebox{0}{\includegraphics*[scale=0.35]{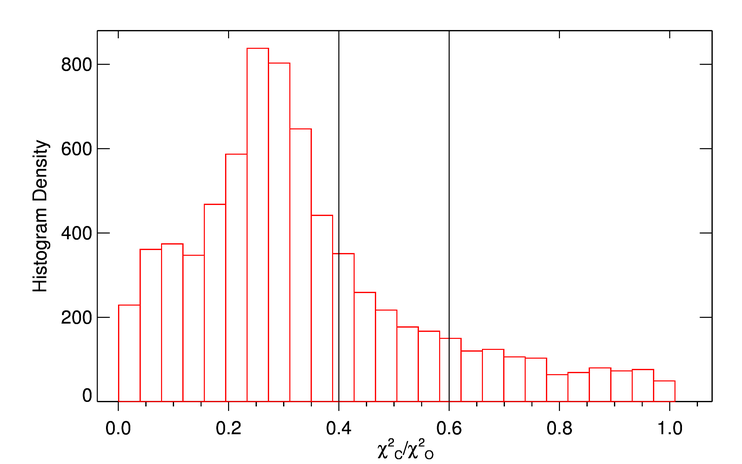}}  
\end{tabular}
\caption{Plot of the ratio $\chi^2_{\rm best}/\chi^2_{\rm alt}$ for the sources classified as O- (top) and C-rich (bottom).  Vertical lines denote the divisions between high, medium, and low confidence sources.}
\label{fig:conf_plot}
\end{center}
\end{figure}

We classify a source as O- or C-rich based on the best-fitting GRAMS model.  We assign a confidence in this classification by examining the ratio $\chi^2_{\rm best}/\chi^2_{\rm alt}$, where the two values of $\chi^2$ are the best fitting models from each GRAMS grid.  Because the $\chi^2$ distributions of the two classes of star peak at different values (Figure~\ref{fig:chi}), we use different criteria to define the confidence intervals for each class.  For sources classified as O-rich, we consider them to be high confidence if $\chi^2_{O}/\chi^2_{C} < 0.1$ and low confidence if $\chi^2_{O}/\chi^2_{C} \ge 0.2$ (medium confidence is intermediate to these two).  For sources classified as C-rich, high confidence is considered to be $\chi^2_{C}/\chi^2_{O} < 0.4$ and the classification is considered low confidence if $\chi^2_{C}/\chi^2_{O} \ge 0.6$.  Table~\ref{tab:conf} gives the number of stars in each category.  Figure~\ref{fig:conf_plot} shows a histogram of these ratios for both categories of sources.  The vertical lines denote the points dividing high, medium and low confidence.  The division points were selected to fall where the source density has fallen by a factor of approximately 2.

We have tested our classification against the spectroscopically verified sources used by \citet{Groenewegen2009} and \citet{vanLoon1999}.  All sources were matched to our catalog using a 2\arcsec\ matching radius.  Of the 66 C-rich sources used in \citet{Groenewegen2009} that we find in our sample, we correctly classify 60 as C-rich.  Of 27 C-rich sources in \citet{vanLoon1999}, we classify 24 of them as C-rich as well.  These numbers correspond to 91\% and 89\% agreement, respectively.  The O-rich sources show similarly good agreement.  We find 40 of the sources that \citet{Groenewegen2009} classify as O-rich in our sample, 38 of which we also identify as O-rich.  Of the O-rich sources identified by \citet{vanLoon1999}, 14 are found in our sample and 10 are identified by GRAMS as O-rich.

In addition, we compared our sample to the surveys of \citet{Blanco1980} and \citet{Blanco1990}.  We identify 177 carbon stars from \citet{Blanco1980} in our sample, correctly classifying 145 (82\%) of them.  Of the 96 O-rich stars from \citet{Blanco1980} identified in our dataset, GRAMS classified 92 (96\%) correctly.  \Citet{Blanco1990} focused exclusively on C-rich stars, and we find 538 of their sources in our list, with 426 (79\%) classified correctly by us.

We also compare our sample to the point sources from the SAGE-Spec program \citep{Kemper2010} classified by \citet{Woods2011}.  We find 87 sources from that project with valid GRAMS model fits in our own data, and excellent agreement between the two classifications.  Of 17 SAGE-Spec RSGs, 100\% are matched to GRAMS O-rich models, of which the RSG models are a subset.  Of the other 70 SAGE-Spec sources, only 3 have conflicting GRAMS and SAGE-Spec classifications, a 96\% success rate.  Two of the 3 sources with conflicting classifications are low-confidence as defined earlier in this section (Figure~\ref{fig:mis}, top two panels).  The remaining source (SSTISAGEMA J053027.49-690358.3, Figure~\ref{fig:mis}, bottom panel) is technically a high confidence source, but examination of its SED shows that it is extremely well fit by both an O-rich ($\chi^2_{O}=1.5$) and a C-rich model ($\chi^2_{C}=5.9$).  The fits are essentially indistinguishable for this source.  Combining all the above comparisons, we conclude that GRAMS has a greater than 80\% accuracy rate when compared to spectroscopic classifications, correctly classifying 786 out of 948 spectroscopically classified stars to which our results were compared.  The sources which are misclassified by GRAMS lie very close to the O- and C-rich dividing line from \citet{Cioni2006} in the \ks\ vs. \jmag\ $-$ \ks\ CMD, squarely atop the locus of low-confidence classifications (green points, Figure~\ref{fig:conf_plot}).

The 12 spectroscopically confirmed carbon stars from \citet{Gruendl2008} are all correctly identified as C-rich by GRAMS.  We find that the current version the GRAMS grid does not contain models of sufficient optical depth to reproduce the 24 and 70~\mic\ photometry given in that work, and we therefore use the luminosities and dust mass-loss rates derived by those authors.  Beyond noting in passing that GRAMS did identify them as C-rich, these sources are only included in our dataset for discussions of the global dust-mass injection into the ISM of the LMC.  It should be noted that the  mass-loss rates obtained by \citet{Gruendl2008} were derived using the same set of optical constants used by \citet{Groenewegen2007} which, as discussed in more detail in \citetalias{Srinivasan2011}, yield values of \mlr\ systematically 2-4 times higher than those produced by a GRAMS model of identical optical depth.

\begin{deluxetable*}{lrrrr}
\tabletypesize{\scriptsize}
\tablewidth{0pt}
\tablecaption{Confidence Intervals}
\tablecomments{Number of classifications considered to be `high,' `medium,' or `low' confidence.  Confidence intervals are defined based on the ratio $\chi^2_{\rm best}/\chi^2_{\rm alt}$, where the two values of $\chi^2$ are those of the best fitting model, and the best fitting model from the other GRAMS grid (O- or C-rich).  See text for details.}
\tablehead{ \colhead{Class (GRAMS)} & \colhead{Total} & \colhead{High Confidence} & \colhead{Medium Confidence} & \colhead{Low Confidence} }
\startdata
O-rich AGB &        26\,210 &        16\,609 (63.4\%) &         4917 (18.8\%) &         4684 (17.9\%) \\
C-rich AGB &         7281 &         5213 (71.6\%) &         1116 (15.3\%) &          952 (13.1\%)
\enddata
\label{tab:conf}
\end{deluxetable*}

\begin{figure}
\begin{center}
\begin{tabular}{c}
\rotatebox{0}{\includegraphics*[scale=0.35]{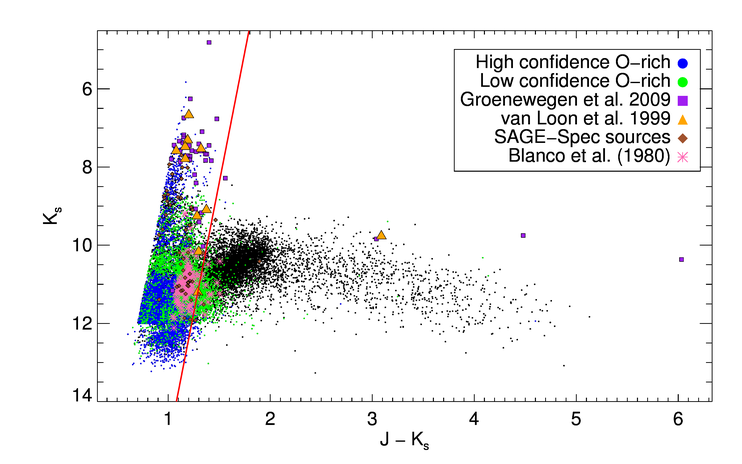}} \\
\rotatebox{0}{\includegraphics*[scale=0.35]{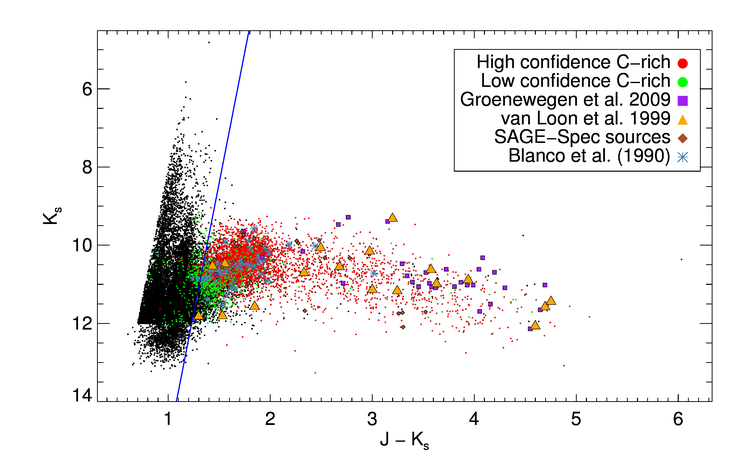}}  
\end{tabular}
\caption{\ks\ vs. \jmag\ $-$ \ks\ CMD showing the sources classified as O-rich (top) or C-rich (bottom) sources by GRAMS with high (O-rich: blue, C-rich: red) or low (green, both panels) confidence.  Both panels feature the CMD-based division line proposed by \citet{Cioni2006}.  For both classifications of star, the low confidence source are clustered near the CMD-based division.  Of the 1337 sources with different CMD and GRAMS classifications, 1163 are identified as low confidence.}
\label{fig:conf_cmd}
\end{center}
\end{figure}

\begin{figure}
\begin{center}
\begin{tabular}{c}
\rotatebox{0}{\includegraphics*[scale=0.35]{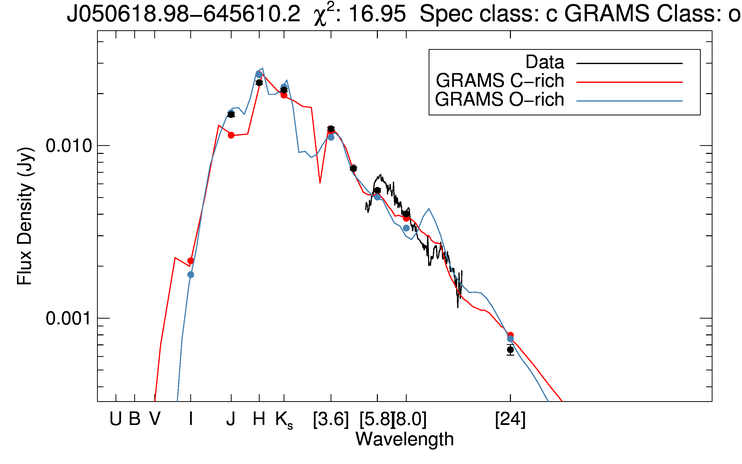}} \\
\rotatebox{0}{\includegraphics*[scale=0.35]{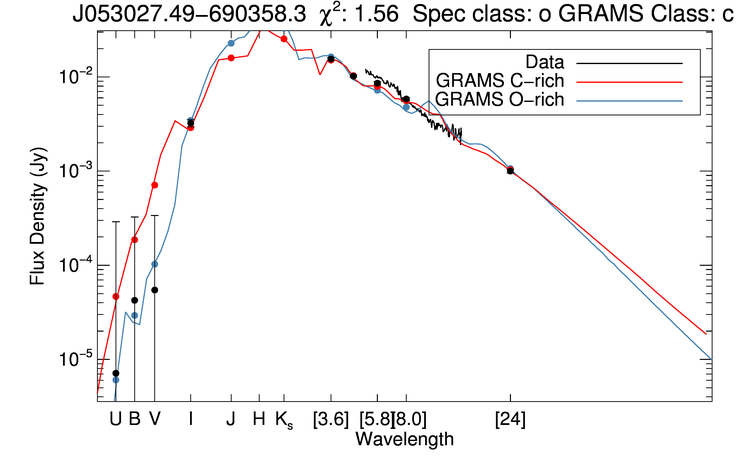}} \\
\rotatebox{0}{\includegraphics*[scale=0.35]{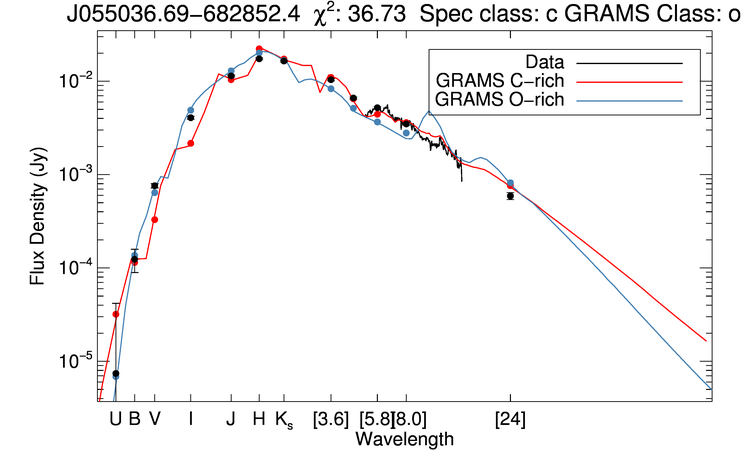}}
\end{tabular}
\caption{The 3 sources in our sample with different SAGE-Spec and GRAMS classifications.  \textit{Top/Middle Panels}: These sources are classified by the GRAMS grid as low-confidence O-rich sources, a differing spectral classification is not surprising.  \textit{Bottom Panel}: This source is technically a high-confidence classification (\S~\ref{sec:oc}), but is fit excellently by both an O- and C-rich model.  Again, a conflict between the spectrum and the RT model is not troubling.}\label{fig:mis}
\end{center}
\end{figure}

Figure~\ref{fig:conf_cmd} presents a comparison between the CMD-based classification scheme proposed by \citet{Cioni2006} and our current results.  The top panel of the figure presents the sources we classify as O-rich, with high confidence sources in blue, and low confidence sources in green.  The bottom panel focuses on C-rich sources, using red for the high confidence sources and green again for low confidence sources.  Both panels show the line defined by \citet{Cioni2006} which divides O-rich AGBs from C-rich AGBs in their schema.  The GRAMS classification is in very good agreement with the CMD-based classification.  As the line from \citet{Cioni2006} is based on only two bands of photometry, and GRAMS is based on 12, we interpret this agreement as giving support to the CMD classification scheme.  The fact that out of the 1337 sources in this CMD with differing CMD/GRAMS classifications, 1163 (87\%) are considered low confidence classifications lends support to our established confidence intervals.  Where the GRAMS classification disagrees with that derived from the \citet{Cioni2006} criteria, we give preference to the GRAMS classification.

\subsection{Luminosity Functions} \label{sec:lum_func}
\begin{figure}
\begin{center}
\rotatebox{0}{\includegraphics*[scale=0.35]{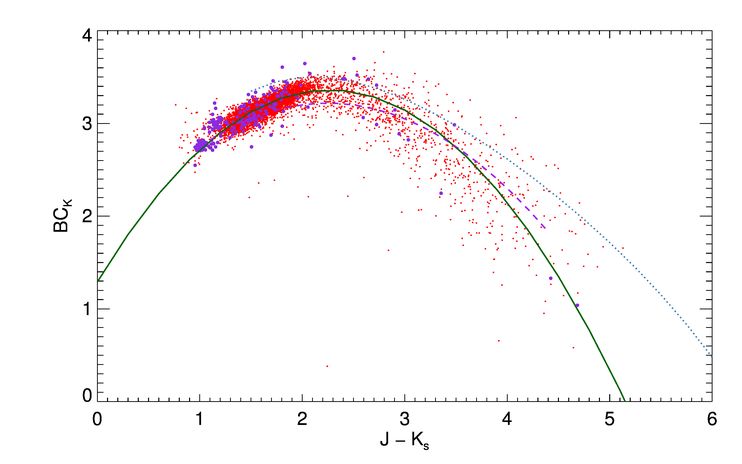}}
\caption{Bolometric correction for C-rich AGB candidates at \ks\ as a function of \jmag\ $-$ \ks.  The C-rich sample from this work is shown in red, and the best-fit second-order polynomial to the data is shown in green.  For comparison, the C-rich sources from \citet{Marco2012} are shown in as purple points.  The fit is to all points in the figure.  The dotted curve is the $4^{\rm th}$ order polynomial derived in \citet{Whitelock2006}.  The violet dashed curve is the relation derived by \citet{Kerschbaum2010}.}
\label{fig:bc}
\end{center}
\end{figure}

By covering the spectral region 0.2--1000~\mic, the GRAMS grid provides accurate determinations of the bolometric luminosity of our sources.  With this well-determined bolometric luminosity, we derive an expression for the bolometric correction at \ks\ (BC$_{\rm K_{s}}$) as a function of \jmag\ $-$ \ks\ color.  We find a second-order polynomial fits our data well (Figure~\ref{fig:bc}).  Also included in the figure is the quartic fit proposed by \citet{Whitelock2006}, and the quadratic fit obtained by \citet{Kerschbaum2010}.  Our fit is of the form $BC_{K_s} = a_0 + a_{1}(J-K_s) + a_{2}(J-K_s)^2$ and is detailed in Table~\ref{tab:bc_fit}.  We overlay the C-rich sample from \citet{Marco2012} as blue points atop our our dataset shown in red.  We note the remarkable consistency between these two samples.

\begin{deluxetable}{lc}
\tabletypesize{\scriptsize}
\tablewidth{0pt}
\tablecaption{BC$_{K_s}$ vs.~$J-K_s$ Quadratic Fit}
\tablecomments{Fit parameters for the best fit quadratic function to BC$_{K_s}$ as a function of $J-K_s$ color.  N is the number of sources used in calculating the fit, and $\sigma$ is the MAD of the residuals to the fit.}
\tablehead{\colhead{Quantity} & \colhead{Value}}
\startdata
a$_0$ &  1.29 \\
a$_1$ &  1.83 \\
a$_2$ &  -0.40 \\
N &  \numc \\
$\sigma$ &    0.06
\enddata
\label{tab:bc_fit}
\end{deluxetable}

\begin{figure}
\begin{center}
\rotatebox{0}{\includegraphics*[scale=0.35]{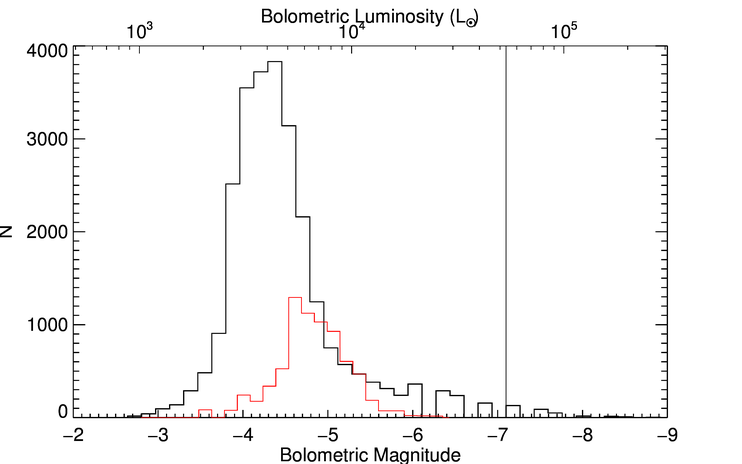}}
\caption{Luminosity functions for the stars in our sample.  O-rich candidates are in black, C-rich candidates in red.  The vertical line represents the classical AGB luminosity limit, $\sim$54\,000~L$_{\odot}$.}
\label{fig:lum_func}
\end{center}
\end{figure}

Figure~\ref{fig:lum_func} shows the luminosity function (LF) for the O-rich (black) and C-rich (red) populations of the LMC.  RSGs are included in the O-rich sample.  The maximum luminosity for an AGB star \citep[$\sim$54\,000~L$_{\odot}$][]{Paczyski1971} is indicated with a vertical black line.  The `picket fence' effect seen at high luminosities is an artifact of increasingly coarse grid coverage at these high luminosities (see also Figure~\ref{fig:grams_cmd}), not a true characteristic of the RSG population luminosity function.  The brightest O-rich source in our sample is a RSG far in excess of the maximum AGB luminosity limit, with a luminosity of 1\,000\,000~L$_{\odot}$.  The brightest C-rich AGB star in our sample is 25\,700~L$_{\odot}$, well below the maximum luminosity for a shell hydrogen/helium burning source.

The O-rich distribution is concentrated at a lower luminosity than the C-rich distribution.  C-rich sources are generally more evolved than O-rich sources, having had time to experience several thermal pulses and dredge up sufficient carbon to become C-rich.  As luminosity increases with age along the AGB \citep[][though it should be noted that this is not a monotonic or smooth increase, but more of a `two steps forward, one step back' process]{Vassiliadis1993}, C-rich AGB stars should be brighter than O-rich AGB stars (see also Fig~\ref{fig:per_mag}).  However, the O-rich distribution does extend a tail to far higher luminosities than the C-rich population reaches.  The most massive AGB stars ignite the bottom of their convective hydrogen envelopes, and this `hot-bottom burning' can make them burn brighter than the classical AGB limit \citep{Bloecker1991}.  Additionally, the AGB limit, obviously, does not apply to core-helium burning RSGs, which are also included amongst our O-rich grid.  These two populations extend the bright tail of the O-rich LF, while the median of the C-rich LF remains brighter than that of the O-rich population.

\begin{figure}
\begin{center}
\rotatebox{0}{\includegraphics*[scale=0.35]{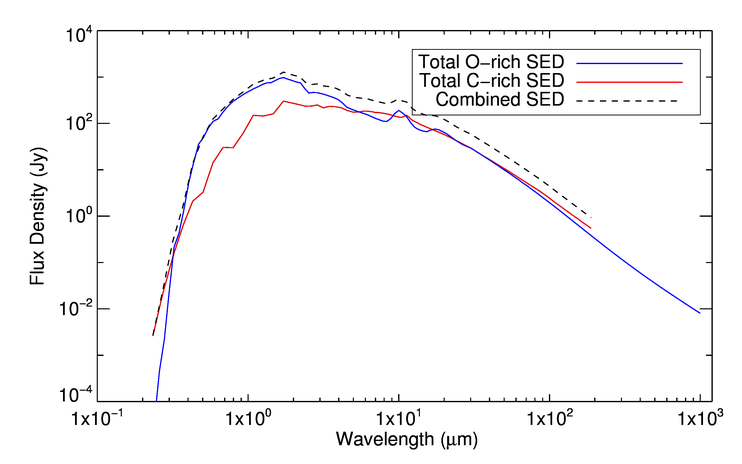}}
\caption{Plot of the integrated SED for the entire sample.  Red curve: sum of all used GRAMS C-rich model SEDs, blue curve: sum of all used GRAMS O-rich SEDs, black curve: sum of red and blue curves.  This figure constitutes a prediction for the integrated SED of the evolved stellar population of unresolved galaxies with metallicity and star formation history similar to that of the LMC.}
\label{fig:total_flux}
\end{center}
\end{figure}

Figure~\ref{fig:total_flux} shows the integrated SED of the O- and C-rich populations of our sample individually, together with the combined total of our entire dataset.  The figure was produced by simply summing the SED from each star's associated best-fit GRAMS model.  This simulates the expected observed SED of the evolved stellar population of the entire LMC if it were unresolved.  We note that the O-rich population (and its RSG sub-population) dominate the SED of the entire population at near-IR wavelengths around 1~\mic.   The lesser number of C-rich stars means that they do not outshine the O-rich population (except by a small amount at mid-IR wavelengths), but do serve to ``wash-out" the silicate feature at $\sim$9~\mic, making it much less prominent than if the O-rich stars were observed in isolation.

\subsection{Distribution of Dust Mass-Loss Rate Along the AGB} \label{sec:mlr}
One of the primary motivators for the study of AGB stars is their extensive mass loss.  The broad picture of this process is understood \citep[see, e.g.][]{Mattsson2011}: hydrodynamic pulsations in the interiors of AGB stars develop, due to the steep density gradients in the extremely extended atmospheres of these stars, into shocks which eject the outer layers of the atmosphere.  When this ejected material cools below the dust condensation temperature ($\sim$1000~K), dust forms.  Due to its greater opacity, this dust is accelerated by the radiation pressure from the star, and drags along the gas to which it is collisionally coupled.  While it makes up only $\sim$1\% by mass of the material ejected from AGB stars, the dust takes on disproportionate scientific importance, due to its role as the driver of mass loss and because it is more easily observed in the infrared than the gas.  The dust mass-loss rate is one of the stellar parameters spanned by the GRAMS model grid,  and we have determined the \mlr\ for all $\sim$30\,000 sources in our sample.  Throughout this work, we report only the dust mass-loss rate, not multiplying by a gas to dust ratio, $\psi$.  Our reported mass-loss rates will thus be about 2 orders of magnitude smaller than values reported by studies which report total (gas and dust) mass loss \citep[e.g][]{Groenewegen2009}.

\begin{figure}
\begin{center}
\rotatebox{0}{\includegraphics*[scale=0.32]{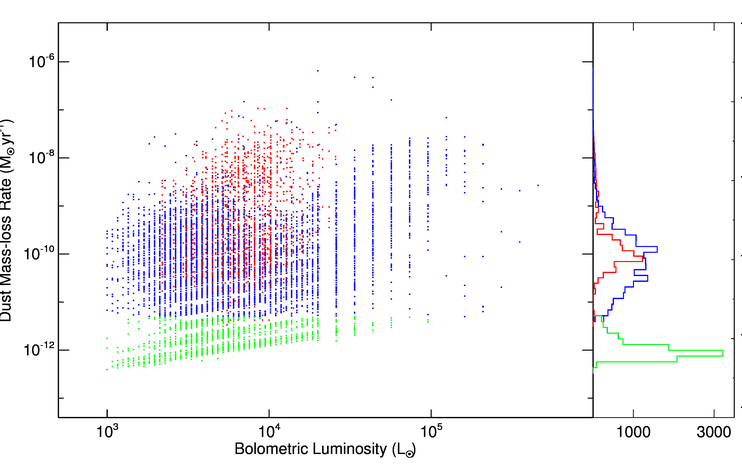}}
\caption{\mlr\ vs.\ bolometric luminosity for our sample.  O-rich sources plotted in blue, C-rich sources in red.  The left side of this figure is a scatter plot of all 30\,000 sources in our sample.  The discrete nature of the GRAMS grid leads to sources having best-fit models with identical values of \mlr\ and luminosity, these sources are plotted atop one another in this figure.  The left panel shows no clear trend of \mlr\ with luminosity.  The right hand panel of this figure is a histogram of \mlr\ using the same color coding  and scale as in the scatter plot.  The O-rich sources are clearly bi-modal, and the minimum in this histogram ($\log(\dot{M_{d}}) \approx -11.3$) serves as the division between the low-\mlr\ sources and the rest of the O-rich population.}
\label{fig:mlr_lum}
\end{center}
\end{figure}

Figure~\ref{fig:mlr_lum} presents a plot of \mlr\ vs.\ bolometric luminosity, with a histogram of \mlr\ presented along the right side y-axis.  Luminosity and \mlr\ are presented on logarithmic axes, the histogram is linear.  As both \mlr\ and luminosity are outputs of the GRAMS grid, the discrete nature of the grid is visible in this plot, particularly at the high luminosities.  The fact that the grid is discrete also results in numerous ``collisions," where multiple stars are fit with models with identical values of \mlr\ and luminosity.  These stars are plotted atop each other in the scatter plot, and are not visible as separate points.  There is a slight tendency for \mlr\ to increase with increasing luminosity, but no quantifiable trend.  For a given star, there is expected to be a trend of increasing \mlr\ with increasing luminosity during the course of its evolution.  Since our sample consists of stars of various initial masses, various states of evolution, and various points along the thermal and hydrodynamical pulse cycles, this relation is washed out by the numerous complicating factors of AGB evolution.  The \mlr\ histogram along the y-axis shows that C-rich stars are clustered at the high end of the mass-loss range, and the O-rich stars are extremely bimodal.  We identify a sub-population of O-rich sources based on this bi-modality, using \mlr\ greater or less than $\log(\dot{M_{d}}) = -11.3$ as the dividing line.  There are \numlo\ sources in the low \mlr\ population and \numhi\ in the high \mlr\ group.  The GRAMS grid was intentionally designed to cover a wider range of stellar and circumstellar shell properties than are spanned by actual AGB stars, and while the best fit models for these sources do have non-zero mass-loss rates, the rates are so low as to be consistent with zero, and we do not consider these low mass-loss rates to necessarily be physical.

\begin{figure}
\begin{center}
\rotatebox{0}{\includegraphics*[scale=0.32]{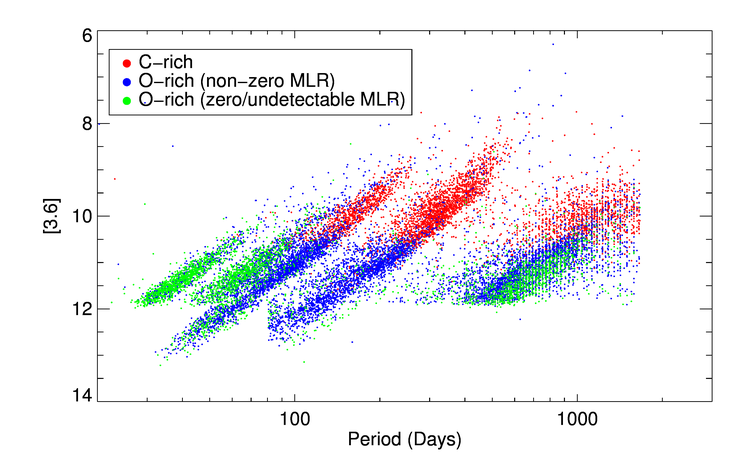}}
\caption{Plot of [3.6] magnitude vs.\ variability period.  C-rich candidates in red, O-rich candidates with $\log(\dot{M_{d}}) < -11.3$ in green, all other O-rich sources in blue.  The 5 parallel AGB Period-luminosity sequences are visible (see also Figure~2, \citet{Riebel2010}).  Low-\mlr\ sources, thought to be the least evolved sources, are clustered on sequences 3, 4 and D.}
\label{fig:per_mag}
\end{center}
\end{figure}

We highlight these low-\mlr\ sources in Figure~\ref{fig:per_mag}.  This plot shows the [3.6]~magnitude vs.\ the $\log$(Variability Period).  C-rich candidates are plotted in red, O-rich sources with $\log(\dot{M_{d}}) \ge -11.3$ are in blue, and the low-\mlr\ O-rich sources are highlighted in green.  Visible in this figure are the 5 parallel AGB pulsation sequences, labeled 4--1 and D from left to right \citep[Figure~1 from][]{Riebel2010}.  As discussed in \citet{Vassiliadis1993}, stars are expected to begin their evolution along the AGB near the bottom of sequence 4, and generally evolve to longer periods and to brighter luminosities, finishing their lives pulsating in the fundamental mode (sequence 1).  The mechanism behind sequence D, the `long secondary period,' is currently unknown \citep{Nicholls2009,Wood2009}.  We would expect the least-evolved AGB stars to exhibit the lowest rates of mass-loss, and we see that the low-\mlr\ sources defined above cluster at the bottom of sequences 3 and 4, giving us confidence that these sources indeed represent the least-evolved sources in our sample, consistent with bare photospheres.  Indeed, the low-\mlr\ population is clustered almost exclusively on sequences 3 \&\ 4, with almost none on sequences 2 \&\ 1.  There is essentially no such thing as a non-mass-losing AGB star on the fundamental-mode sequence.  We note, however, that the half of our sample with well-determined MACHO periods (that appear in Figure~\ref{fig:per_mag}) constitutes only 20\% of the total \mlr\ of our entire sample.  A full 80\% of the total dust injection from evolved stars into the ISM comes from sources too red to have been detected by the MACHO survey, lack quality variability information, and therefore are not included in this figure.  In a future paper, we will present initial constraints on the variability of these sources, using the warm \textit{Spitzer} mission.

\begin{deluxetable}{lrr}
\tabletypesize{\scriptsize}
\tablewidth{0pt}
\tablecaption{Sequence D Comparison}
\tablecomments{Comparison of the population of Sequence D with the radial pulsation P-L sequences 1--4 \citep[see][]{Riebel2010}.  The O-rich AGB stars in sequence D have the same ratio of high to low \mlr\ as the population of the other 4 sequences, consistent with sequence D population being drawn randomly from sequences 1--4.  However, Sequence D is noticeably poor in sources classified as C-rich as compared to the other sequences. }
\tablehead{ \colhead{Classification} & \colhead{Sequences 1--4} & \colhead{Sequence D} }
\startdata
C-rich                                 &  2906 (26\%) & 845 (19\%) \\
O-rich (all)                           &  7894 (73\%) & 3505 (80\%) \\
O-rich ($\log(\dot{M_{d}}) \ge -11.3$) &  5405 (68\%) & 2438 (69\%) \\
O-rich ($\log(\dot{M_{d}}) < -11.3$)   &  2496 (31\%) & 1067 (30\%) \\
All Sources                            &  10\,800     & 4350
\enddata
\label{tab:seqd}
\end{deluxetable}

There is also a population of low-\mlr\ sources on Sequence D.  Table~\ref{tab:seqd} summarizes the relative numbers of each type of source on sequence D as compared to the other sequences \textit{in toto}.  Sequence D has the same \mlr\ population distribution as Sequences 1--4 ($\sim$1/3 low-\mlr).  Sequence D is more O-rich than the radial pulsation sequences.  The C- to O-rich ratio among the sequence D sources is only 24\%, compared to 37\% in the radially pulsating sequences.  Bootstrap analysis shows that the difference in C- to O-rich ratio between sequence D and the other sequences is not consistent with sequence D being a randomly drawn subsample of the radially pulsating sequences.

\subsection{The Total Dust Mass Return to the ISM} \label{sec:int_mlr}

\begin{figure*}
\begin{center}
\begin{tabular}{cc}
\rotatebox{0}{\includegraphics*[scale=0.3]{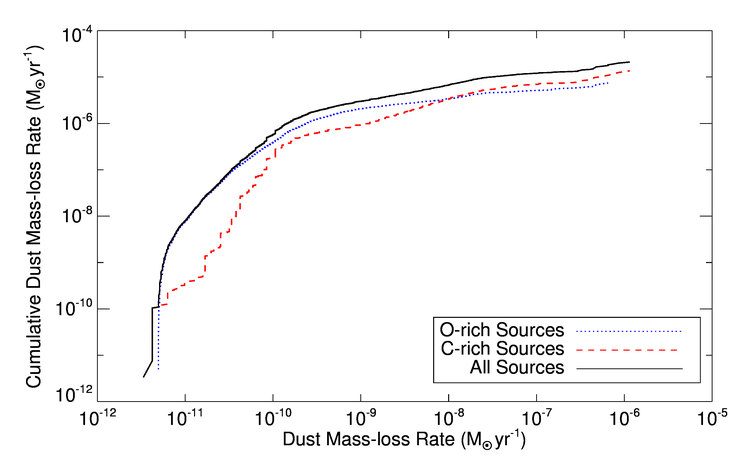}} & \rotatebox{0}{\includegraphics*[scale=0.3]{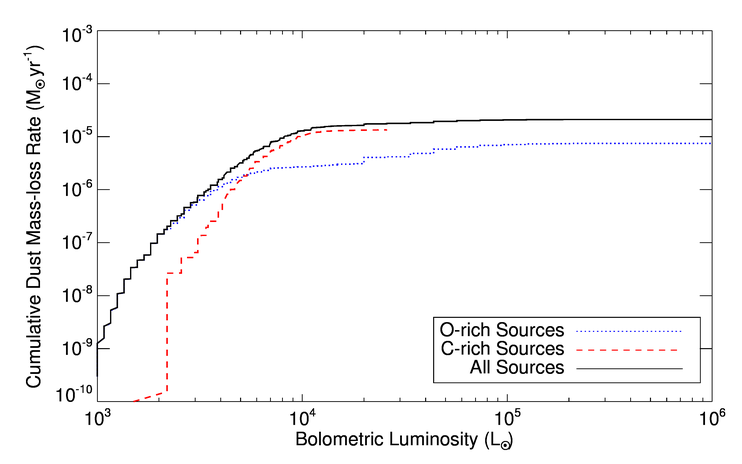}}
\end{tabular}
\caption{\textit{Left Panel}: Cumulative \mlr\ vs.\ \mlr\ for individual sources.  Dashed line: C-rich sources, dotted line: O-rich sources, solid line: total of all sources.  The O-rich sources dominate the lower \mlr\ region through their larger population.  The C-rich sources are less numerous, but dominate at the largest values of \mlr.  \textit{Right Panel}: Cumulative \mlr\ vs.\ bolometric luminosity of individual sources.  The same linestyles are used as in the left panel.  Here O-rich sources, which contain both hot-bottom burning stars and red supergiants, extend to much higher luminosities, but it is the C-rich sources that dominate much more strongly.}
\label{fig:cum_mlr}
\end{center}
\end{figure*}

Figure~\ref{fig:cum_mlr} shows the cumulative mass return to the ISM from AGB and RSG stars, as a function of \mlr\ (left panel) and luminosity (right panel).  Both panels show the running total for both C- and O-rich sources separately, and for both combined.  The O-rich grid extends to both higher luminosities (because it is intended to cover RSGs) and higher values of \mlr\ than the C-rich grid, but the C-rich stars contribute greatly to the total dust injection to the ISM, accounting for over half the dust production but only 21\% of the sample by number.  The total AGB+RSG dust mass return, $(2.1 \pm 0.06) \times 10^{-5}$~\msunperyr is dominated by a very small fraction of stars.  In agreement with \citetalias{Srinivasan2009} and \citet{Boyer2011}, we find that the total contribution of dust to the ISM from AGB stars is dominated by a very few `extreme' stars.  There are \numx\ stars in our sample classified as `extreme' using CMD cuts.  Of these, \numxc\ (97\%) are classified as C-rich by the GRAMS grid.  These C-rich `extreme' stars comprise only 4\% of our sample by number, but account for 75\% of the total \mlr.  The few `extreme' stars classified as O-rich are also heavy mass-losers, making up only 0.1\% of our sample by number but representing 13\% of the total dust-injection rate.  We find a total AGB dust production rate of $(1.91 \pm 0.06) \times 10^{-5}$~\msunperyr. This is quite consistent with the work of \citet{Srinivasan2009}.  Table~\ref{tab:mlr} breaks down our estimates of total \mlr\ for each sub-population in our sample.
We obtain exactly the same value for the dust injection rate due to RSGs as \citet{Matsuura2009}, $2 \times 10^{-6}$\msunperyr.  Our results indicate that the dust production in the LMC is dominated by C-rich AGBs, but not to the same extent as found by \citet{Matsuura2009}.  

Our estimate of the global dust production of the AGB population in the LMC is a factor of $\sim$2 smaller than that of \citet{Matsuura2009}.  This discrepancy may in part be due to the systematically lower dust mass-loss rates in GRAMS carbon-star models than in the models used in that work \citepalias[see \S~5.1.2 in][]{Srinivasan2011}.  Additionally, recent work by \citet{Matsuura2011} indicates that SNe may produce substantially more dust than previously estimated, up to $\sim$1~M$_{\odot}$ each.  If this result is confirmed, our downgrade of the dust contribution of AGB stars serves to indicate that SNe may be the dominant source of dust in the ISM of the LMC.  We would like to note, however, that care should be taken when comparing our results to other studies. For instance, the GRAMS grid is constructed with the assumption of a constant dust shell expansion velocity $v_{\rm exp}$, which is kept fixed at 10~km~s$^{-1}$. The actual value of $v_{\rm exp}$ for LMC stars is uncertain, and it may depend on the luminosity, metallicity and gas:dust ratio \citep[see, e.g.][]{vanLoon2000}. The uncertain value of the expansion velocity translates to an uncertainty in the absolute value of the dust mass-loss rate and, therefore, the global dust injection rate.

\begin{deluxetable}{lcc}
\tabletypesize{\scriptsize}
\tablewidth{0pt}
\tablecaption{Total \mlr\ by population}
\tablecomments{Total of \mlr\ broken down by classification.  Column 3 lists the fraction of the total evolved star dust mass injection to the ISM each population contributes.  Note that the category ``Extreme AGBs" is a subset of O-rich AGB and C-rich AGB (most extremes are C-rich).}
\tablehead{\colhead{Population} & \colhead{Total \mlr\ ($\times 10^{-6}$\msunperyr)} & \colhead{Percent of total} }
\startdata
All Sources & $  21.1 \pm 0.6$ &  100.0\% \\
C-rich AGBs & $ 13.64 \pm 0.62$ &   64.6\% \\
O-rich AGBs & $   5.5 \pm 0.2$ &   26.0\% \\
RSGs & $   2.0 \pm 0.1$ &    9.4\% \\
Extreme AGBs & $  15.7 \pm 0.6$ &   74.2\% 
\enddata
\label{tab:mlr}
\end{deluxetable}

\subsection{Observational Proxies for \mlr} \label{sec:proxy}
Because of the observational investment necessary to obtain 12 bands of photometry for \mlr\ modeling purposes, or precise CO line profiles to trace the gas return to the ISM directly, it is desirable to develop simple, easily observable proxies for this astrophysical quantity.  We fit hyperbolas to \mlr\ as a function of all possible IR colors in our dataset.  Not all colors serve as useful proxies for \mlr\, and we present selected fits in table~\ref{tab:mlr_fit}.  We have fit our functions to only the C-rich AGB population in our sample, as we find that the O-rich AGB and RSGs populations are very condensed in color space, and do not follow a well-defined empirical relationship.  The primary signature of circumstellar dust in O-rich AGB stars is the silicate feature at $\sim$9~\mic, which is not adequately sampled by the IRAC camera aboard \textit{Spizter}.  The [9]~\mic\ band provided by the AKARI satellite is uniquely positioned to probe this spectral feature, we hypothesize that \ks\ $-$ [9] would be an excellent proxy for \mlr\ in O-rich AGB stars, and intend to investigate this in future work.

\begin{deluxetable}{lcccc}
\tabletypesize{\scriptsize}
\tablewidth{0pt}
\tablecaption{C-rich AGB \mlr\ vs.\ IR color}
\tablecomments{Hyperbolic functions of IR colors as observable proxies for \mlr\ in C-rich AGB stars.  Listed are the coefficients of the best fit hyperbola, of the form $\log \dot{M_{d}} = \frac{{\rm P}_0}{({\rm color})+{\rm P}_1}+{\rm P}_2$.  The `residual' column lists the MAD of the residuals to the best-fit hyperbola, a measure of the spread of the data about the best fit curve.}
\tablehead{\colhead{Color} & \colhead{P$_0$} & \colhead{P$_1$} & \colhead{P$_2$} & \colhead{Residual} }
\startdata
J -- H &     -6.32 &      0.56 &     -6.01 &      0.15 \\
H -- [4.5] &    -24.50 &      4.52 &     -5.67 &      0.09 \\
H --  [8.0] &    -28.18 &      4.89 &     -5.70 &      0.08 \\
H -- [24] &    -29.43 &      5.24 &     -5.88 &      0.09 \\
\ks\ -- [3.6] &     -9.72 &      1.91 &     -6.10 &      0.11 \\
\ks\ -- [4.5] &    -13.94 &      2.88 &     -5.97 &      0.12 \\
\ks\ -- [5.8] &    -17.90 &      3.64 &     -5.88 &      0.15 \\
\ks\ --  [8.0] &    -15.40 &      2.95 &     -6.16 &      0.09 \\
\ks\ -- [24] &    -18.33 &      3.59 &     -6.25 &      0.12 \\
{[}3.6] --  [8.0] &     -9.37 &      1.79 &     -5.85 &      0.15 \\
\enddata
\label{tab:mlr_fit}
\end{deluxetable}

\begin{figure*}
\begin{tabular}{cc}
\rotatebox{0}{\includegraphics*[scale=0.3]{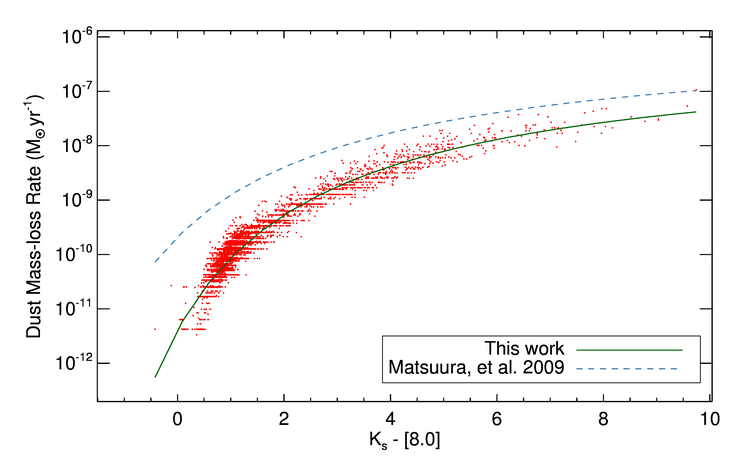}} & \rotatebox{0}{\includegraphics*[scale=0.3]{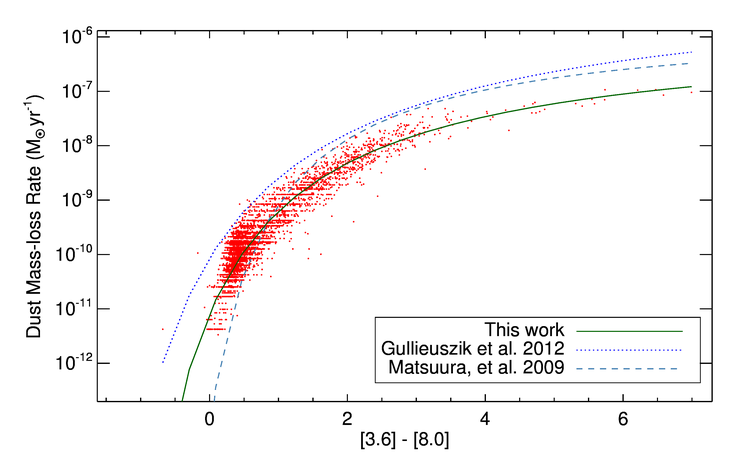}}
\end{tabular}
\caption{\textit{Left panel}: Plot of \mlr\ vs. \ks\ $-$ [8.0] color for C-rich sources.  The solid curve in green is the best-fit line to the C-rich stars: $\log \dot{M_{d}} = \frac{-15.40}{(K_{s} - [8.0])+2.95}-6.16$.  The dashed line is the relation used by \citet{Matsuura2009} (see text for details).  \textit{Right panel}: Plot of \mlr\ vs. [3.6] $-$ [8.0] color for C-rich sources.  The solid curve in green is the best-fit line to the C-rich stars: $\log \dot{M_{d}} = \frac{-9.37}{(K_{s} - [8.0])+1.79}-5.85$.  The dashed curve is the relation used by \citet{Matsuura2009}, and the dotted curve is that derived by \citet{Marco2012} (see text for details).}
\label{fig:mlr_col}
\end{figure*}

Figure~\ref{fig:mlr_col} shows the relationship between \mlr\ and two colors (\ks\ $-$ [8.0] and [3.6] $-$ [8.0]), derived from our sample.  We have chosen these colors for easy comparison to other studies, namely \citet{Matsuura2009} and \citet{Marco2012}.  In order to derive the combined dust budget for the LMC, \citet{Matsuura2009} used two-band color proxies for \mlr\ developed by \citet{Groenewegen2007}.  The left panel of Figure~\ref{fig:mlr_col} compares the \ks\ $-$ [8.0] color proxy used in that work (dashed blue line) to the best fit relation to our data (solid green line).  Our relation is detailed in Table~\ref{tab:mlr_fit}.  The right panel of Figure~\ref{fig:mlr_col} shows another example of an IR color as a proxy for \mlr\ from C-rich AGB stars.  Again the solid green line is the best fit to our data, detailed in Table~\ref{tab:mlr_fit}.  The blue dashed curve is the proxy used by \citet{Matsuura2009}, and the dotted blue line is the relation derived by \citet{Marco2012} drawing on the early results of the IR VISTA survey of the LMC.  Note that the curve plotted here has been divided by the gas to dust ratio, $\psi=200$, used by \citet{Marco2012}, who give their results in terms of total gas+dust mass return.  This allows us to compare the mass-loss rates of only dust.  The method used by \citet{Marco2012} is similar to our own, and the agreement between our curve and theirs in encouraging (see also Figure~\ref{fig:bc}).  The offset between the two curves is a result of different optical constants used to model the opacity of amorphous carbon dust used by our two teams.  As our dust is more opaque, a given IR flux implies less total dust in our models compared to theirs, which corresponds to a lower \mlr.  There is currently no clear consensus as to which set of dust opacities is more correct. See \S~5.1.2 in \citetalias{Srinivasan2011} (and references therein) for a thorough discussion of this effect.

\section{CONCLUSIONS}\label{sec:conclusions}
We present the largest sample of evolved stars subjected to radiative transfer modeling to date.  By fitting $\sim$30\,000 stars with $\sim$12 bands of photometry each to pre-computed radiative transfer models from the GRAMS model grid of O- and C-rich AGB stars and RSGs, we compute individual bolometric luminosities and dust mass-loss rates for the entire evolved star population of the LMC.  Our work forms a useful prediction to be tested against models of stellar population synthesis and stellar evolution.

We establish that the GRAMS model grid \citepalias{Sargent2011,Srinivasan2011} is an important new tool for the study of stellar populations.  Capable of generating statistically accurate predictions for difficult-to-observe stellar parameters such as mass-loss rate and luminosity for thousands of stars in a reasonable period of time, GRAMS will be valuable for interpreting the infrared observations of future missions (\S~\ref{sec:fitting}).  Through comparison to previously published studies of AGB stars in the LMC, we show that the GRAMS O- and C-rich identifications are over 80\% accurate (\S~\ref{sec:oc}).

We present a quadratic formula for the \ks\ band bolometric correction of AGB stars as a function of \jmag\ $-$ \ks\ color (\S~\ref{sec:lum_func}).  Our fit is detailed in Table~\ref{tab:bc_fit}.

We characterize the luminosity functions of both O-rich and C-rich AGB stars in the LMC.  The O-rich stars follow a distribution between M$_{\rm bol}$ = -3.5 - -5 centered at $\sim$ -4.3 ($\sim$3500~L$_{\odot}$).  Individual C-rich AGB stars tend to be brighter, distributed between M$_{\rm bol}$ = -4 -- -6~M$_{\rm bol}$, centered at -4.7~M$_{\rm bol}$ ($\sim$5500~L$_{\odot}$).  Despite tending to be less luminous on an individual basis, O-rich sources dominate the integrated IR light from the LMC evolved stellar population, by virtue of their greater numbers (\S~\ref{sec:lum_func}).

We find that the AGB stars exhibiting the `long secondary period' phenomenon, and lying on sequence D are significantly more O-rich than the stellar population on the other, radially pulsating sequences.  We see a qualitative relationship between \mlr\ and pulsation period for radially pulsating AGB stars, in that sources with no detectable mass-loss are concentrated on sequences 3 \&\ 4, and do not appear at all on sequences 1 \&\ 2 (\S~\ref{sec:mlr}).

We also derive hyperbolic fits to the \mlr\ from C-rich AGB stars as a function of various near-IR colors (\S~\ref{sec:proxy}).  Presented in Table~\ref{tab:mlr_fit}, our fits are derived from a sample much larger than those used by previous studies.

We derive a total dust mass-injection rate for the entire RSG+AGB stellar population of ($2.13\pm0.02) \times 10^{-5}$~\msunperyr of the LMC through consistent, direct summation.  Assuming a gas-to-dust ratio $\psi=500$ for O-rich AGB stars and RSGs, and a ratio $\psi=200$ for C-rich AGB stars, this translates to a total mass injection rate into the ISM from RSGs and AGB stars of $\dot{M} = 6.5\times 10^{-3}$~\msunperyr, with half the mass coming from Carbon-rich AGB stars.  Dominated by the uncertainty in the gas-to-dust ratio, this figure is accurate to a factor of $\sim$2.  That is, Carbon stars inject the same amount of \textit{mass} into the ISM as O-rich AGBs, but two and a half times as much \textit{dust}.

\section{Acknowledgments}
This publication makes use of data products from the Two Micron All Sky Survey, which is a joint project of the University of Massachusetts and the Infrared Processing and Analysis Center/California Institute of Technology, funded by the National Aeronautics and Space Administration and the National Science Foundation.  This research was supported by NASA NAG5-12595, and NASA ADP NNX11AB06G.  Conversations with M. Boyer were helpful.  DR acknowledges support from NASA/JPL/Spitzer contract \#1415784.




\begin{thebibliography}{}

\bibitem[Alcock et al.(1999)]{Alcock1999} Alcock, C., et al.\ 1999, \pasp, 
111, 1539 


\bibitem[Blanco et al.(1980)]{Blanco1980} Blanco, V.~M., Blanco, B.~M., 
\& McCarthy, M.~F.\ 1980, \apj, 242, 938 


\bibitem[Blanco 
\& McCarthy(1990)]{Blanco1990} Blanco, V.~M., \& McCarthy, M.~F.\ 1990, \aj, 100, 674 


\bibitem[Bloecker 
\& Schoenberner(1991)]{Bloecker1991} Bloecker, T., \& Schoenberner, D.\ 1991, \aap, 244, L43 


\bibitem[Blum et al.(2006)]{Blum2006} Blum, R.~D., et al.\ 2006, \aj, 132, 
2034 


\bibitem[Boothroyd et al.(1993)]{Boothroyd1993} Boothroyd, A.~I., Sackmann, 
I.-J., \& Ahern, S.~C.\ 1993, \apj, 416, 762 


\bibitem[Boyer et al.(2011)]{Boyer2011} Boyer, M.~L., et al.\ 2011, \aj, 
142, 103 


\bibitem[Cioni et al.(2006)]{Cioni2006} Cioni, M.-R.~L., Girardi, L., 
Marigo, P., \& Habing, H.~J.\ 2006, \aap, 448, 77 


\bibitem[Cioni et al.(2011)]{Cioni2011} Cioni, M.-R.~L., et al.\ 2011, 
\aap, 527, A116 


\bibitem[Fraser et al.(2008)]{Fraser2008} Fraser, O.~J., Hawley, S.~L., 
\& Cook, K.~H.\ 2008, \aj, 136, 1242 


\bibitem[Groenewegen et al.(2007)]{Groenewegen2007} Groenewegen, M.~A.~T., 
et al.\ 2007, \mnras, 376, 313 


\bibitem[Groenewegen et al.(2009)]{Groenewegen2009} Groenewegen, M.~A.~T., 
Sloan, G.~C., Soszy{\'n}ski, I., \& Petersen, E.~A.\ 2009, \aap, 506, 1277 


\bibitem[Gruendl et al.(2008)]{Gruendl2008} Gruendl, R.~A., Chu, Y.-H., 
Seale, J.~P., Matsuura, M., Speck, A.~K., Sloan, G.~C., 
\& Looney, L.~W.\ 2008, \apjl, 688, L9 


\bibitem[Gruendl 
\& Chu(2009)]{Gruendl2009} Gruendl, R.~A., \& Chu, Y.-H.\ 2009, \apjs, 184, 172 


\bibitem[Gullieuszik et al.(2012)]{Marco2012} Gullieuszik, M., et 
al.\ 2012, \aap, 537, A105 

\bibitem[Iben(1983)]{Iben1983} Iben, I., Jr.\ 1983, \apjl, 275, L65 


\bibitem[Ita et al.(2008)]{Ita2008} Ita, Y., et al.\ 2008, \pasj, 60, 435 


\bibitem[Karakas et al.(2010)]{Karakas2010} Karakas, A.~I., Campbell, 
S.~W., \& Stancliffe, R.~J.\ 2010, \apj, 713, 374 


\bibitem[Kemper et al.(2010)]{Kemper2010} Kemper, F., et al.\ 2010, \pasp, 
122, 683 

\bibitem[Kato et al.(2007)]{Kato2007} Kato, D., Nagashima, C.,
Nagayama, T., et al.\ 2007, \pasj, 59, 615


\bibitem[Kerschbaum et al.(2010)]{Kerschbaum2010} Kerschbaum, F., 
Lebzelter, T., \& Mekul, L.\ 2010, \aap, 524, A87 


\bibitem[Matsuura et al.(2009)]{Matsuura2009} Matsuura, M., et al.\ 2009, 
\mnras, 396, 918 


\bibitem[Matsuura et al.(2011)]{Matsuura2011} Matsuura, M., et al.\ 2011, 
Science, 333, 1258 


\bibitem[Mattsson \& H\"ofner(2011)]{Mattsson2011} Mattsson, L., H\"ofner, S.\ 2011, \aap, 533, A42 


\bibitem[Meixner et al.(2006)]{Meixner2006} Meixner, M., et al.\ 2006, \aj, 
132, 2268 


\bibitem[Murakami et al.(2007)]{Murakami2007} Murakami, H., et al.\ 2007, 
\pasj, 59, 369


\bibitem[Ngeow \& Kanbur(2008)]{Ngeow2008} Ngeow, C., \& Kanbur, S.~M.\ 2008, Galaxies in the Local Volume, 317 


\bibitem[Nicholls et al.(2009)]{Nicholls2009} Nicholls, C.~P., Wood, P.~R., 
Cioni, M.-R.~L., \& Soszy{\'n}ski, I.\ 2009, \mnras, 399, 2063 


\bibitem[Paczy{\'n}ski(1971)]{Paczyski1971} Paczy{\'n}ski, B.\ 1971, 
\actaa, 21, 417 


\bibitem[Riebel et al.(2010)]{Riebel2010} Riebel, D., Meixner, M., Fraser, 
O., Srinivasan, S., Cook, K., \& Vijh, U.\ 2010, \apj, 723, 1195 


\bibitem[Reid 
\& Goldston(2002)]{Reid2002} Reid, M.~J., \& Goldston, J.~E.\ 2002, \apj, 568, 931 


\bibitem[Rubele et al.(2011)]{Rubele2011} Rubele, S., Girardi, L., 
Kozhurina-Platais, V., Goudfrooij, P., 
\& Kerber, L.\ 2011, \mnras, 414, 2204 


\bibitem[Sargent et al.(2010)]{Sargent2010} Sargent, B.~A., et al.\ 2010, 
\apj, 716, 878 


\bibitem[Sargent et al.(2011)]{Sargent2011} Sargent, B.~A., Srinivasan, S., 
\& Meixner, M.\ 2011, \apj, 728, 93 


\bibitem[Schaefer(2008)]{Schaefer2008} Schaefer, B.~E.\ 2008, \aj, 135, 112 

\bibitem[Schwarzschild \& H\"arm(1965)]{Schwarzschild1965} Schwarzschild, M., H\"arm, R.\ 1965, \apj, 142, 855 


\bibitem[Skrutskie et al.(2006)]{Skrutskie2006} Skrutskie, M.~F., et al.\ 
2006, \aj, 131, 1163 


\bibitem[Srinivasan et al.(2009)]{Srinivasan2009} Srinivasan, S., et al.\ 
2009, \aj, 137, 4810 


\bibitem[Srinivasan et al.(2010)]{Srinivasan2010} Srinivasan, S., et al.\ 
2010, \aap, 524, A49 


\bibitem[Srinivasan et al.(2011)]{Srinivasan2011} Srinivasan, S., Sargent, 
B.~A., \& Meixner, M.\ 2011, \aap, 532, A54 


\bibitem[Ueta 
\& Meixner(2003)]{Ueta2003} Ueta, T., \& Meixner, M.\ 2003, \apj, 586, 1338 


\bibitem[Uttenthaler 
\& Lebzelter(2010)]{Uttenthaler2010} Uttenthaler, S., \& Lebzelter, T.\ 2010, \aap, 510, A62 


\bibitem[van Loon et al.(1999)]{vanLoon1999} van Loon, J.~T., Groenewegen, 
M.~A.~T., de Koter, A., Trams, N.~R., Waters, L.~B.~F.~M., Zijlstra, A.~A., 
Whitelock, P.~A., \& Loup, C.\ 1999, \aap, 351, 559 

\bibitem[van Loon(2000)]{vanLoon2000} van Loon, J.~T.\ 2000, \aap, 354, 
125 

\bibitem[Vassiliadis 
\& Wood(1993)]{Vassiliadis1993} Vassiliadis, E., \& Wood, P.~R.\ 1993, \apj, 413, 641 


\bibitem[Vijh et al.(2009)]{Vijh2009} Vijh, U.~P., et al.\ 2009, \aj, 137, 
3139 


\bibitem[Wachter et al.(2002)]{Wachter2002} Wachter, A., Schr{\"o}der, 
K.-P., Winters, J.~M., Arndt, T.~U., \& Sedlmayr, E.\ 2002, \aap, 384, 452 


\bibitem[Whitelock et al.(2006)]{Whitelock2006} Whitelock, P.~A., Feast, 
M.~W., Marang, F., \& Groenewegen, M.~A.~T.\ 2006, \mnras, 369, 751 


\bibitem[Whitelock et al.(2008)]{Whitelock2008} Whitelock, P.~A.,
Feast, M.~W., \& van Leeuwen, F.\ 2008, \mnras, 386, 313


\bibitem[Winters et al.(2000)]{Winters2000} Winters, J.~M., Le Bertre, T., 
Jeong, K.~S., Helling, C., \& Sedlmayr, E.\ 2000, \aap, 361, 641 


\bibitem[Wood et al.(1999)]{Wood1999} Wood, P.~R., et al.\ 1999, Asymptotic 
Giant Branch Stars, 191, 151 

\bibitem[Wood 
\& Nicholls(2009)]{Wood2009} Wood, P.~R., \& Nicholls, C.~P.\ 2009, \apj, 707, 573 


\bibitem[Woods et al.(2011)]{Woods2011} Woods, P.~M., et al.\ 2011, \mnras, 
411, 1597 


\bibitem[Zaritsky et al.(2004)]{Zaritsky2004} Zaritsky, D., Harris,
J., Thompson, I.~B., \& Grebel, E.~K.\ 2004, \aj, 128, 1606


\end{thebibliography}
\end{document}